\documentclass[aps,pre,twocolumn,superscriptaddress,floatfix,amsfonts,citeautoscript]{revtex4-1}
\usepackage{times}
\usepackage{graphicx}
\usepackage{dcolumn}
\usepackage{bm}
\usepackage{color}
\usepackage{braket}
\usepackage{float}
\usepackage{amssymb}
\usepackage{amsmath}
\usepackage{multirow}
\usepackage{enumerate}
\usepackage[table, svgnames, dvipsnames]{xcolor}
\usepackage{booktabs}
\usepackage{makecell}
\usepackage{multirow}
\begin{document}

\title{Thermal conductance of one dimensional disordered harmonic chains}

\author{Biswarup Ash}
\affiliation{Department of Condensed Matter Physics, Weizmann Institute of Science, Rehovot, Israel 76100}

\author{Ariel Amir}
\affiliation{School of Engineering and Applied Sciences, Harvard University, Cambridge, Massachusetts 02138, USA}

\author{Yohai Bar-Sinai}
\affiliation{School of Engineering and Applied Sciences, Harvard University, Cambridge, Massachusetts 02138, USA}

\author{Yuval Oreg}
\affiliation{Department of Condensed Matter Physics, Weizmann Institute of Science, Rehovot, Israel 76100}

\author{Yoseph Imry}
\thanks{Deceased}
\affiliation{Department of Condensed Matter Physics, Weizmann Institute of Science, Rehovot, Israel 76100}

\begin{abstract}
We study heat conduction mediated by longitudinal phonons in one dimensional disordered harmonic chains. Using scaling properties of the phonon density of states and localization in disordered systems, we find non-trivial scaling of the thermal conductance with the system size. Our findings are corroborated by extensive numerical analysis. We show that a system with strong disorder, characterized by a `heavy-tailed' probability distribution, and with large impedance mismatch between the bath and the system satisfies Fourier's law.  We identify a dimensionless scaling parameter, related to the temperature scale and the localization length of the phonons, through which the thermal conductance for different models of disorder and different temperatures follows a universal behavior.
\end{abstract}

\maketitle

{\it Introduction-} The study of heat transport via phonons in low dimensional (spatial dimension $d<3$) classical and quantum mechanical systems has attracted considerable theoretical and experimental attention in recent years~\cite{PhysRevLett.119.110602, Cividini_2017, XU2016113, ADhar_review, BONETTO, PhysRevLett.91.044301,Monasterio,RamaswamyPhysRevLett, NarayanPhysRevE,DharPhysRevLett2001,Roya}. 
One of the main objectives of these studies is to understand the scaling of heat flux $J$ which, according to Fourier's law~\cite{ADhar_review}, should scale with the system size $L$ as $J \propto L^{-1}$ ($L$ is measured along the direction of heat propagation). But extensive numerical and analytical studies in the past few decades have revealed the possible violation of Fourier's law for low dimensional systems~\cite{Casher_Lebowitz, Rubin_Greer, Casati_PhysRevLett, DharPhysRevLett2001,LEPRI20031,Hurtado_Pablo}. These studies show that $J \propto L^{{(\gamma-1)}}$ with $\gamma \neq 0$ which in turn implies $L$-dependent thermal conductivity, $\kappa = \lim\limits_{L \to \infty }\lim\limits_{\Delta T\to 0 } \frac{J L}{\Delta T} \propto L^{\gamma}$ ($\Delta T$ being the temperature difference across the system)~\cite{ADhar_review}. The violation of Fourier's law in low-dimensional systems is also observed experimentally in the case of carbon nanotubes~\cite{ChangPRL}, nanowires~\cite{YANG201085} and graphene~\cite{Xu2014}. 

For systems of finite size, instead of thermal \textit{conductivity} $\kappa$, it is useful to study thermal \textit{conductance}, $G = \kappa L^{d-2}$. Thus, for one dimensional systems $(d=1)$, according to Fourier's law, we expect $G(L) \propto L^{-\beta}$, with $\beta=1$ for normal heat transport, while $\beta \neq 1$ implies anomalous heat transport. Note that the scaling exponents $\gamma$ and $\beta$ characterizing the thermal conductivity and thermal conductance, respectively, are related by $\beta=1-\gamma$. One interesting question is under what conditions $\beta=1$ (or, $\gamma=0$)? 

Various aspects, such as disorder~\cite{Casher_Lebowitz,Herrera_Gonz_lez_2015,Ariel_Amir}, phonon-phonon interaction~\cite{PhysRevLett.100.134301,PhysRevE.57.2992}, presence of pinning potential~\cite{PhysRevE.78.051112,PhysRevE.87.020101}, nature of the heat baths~\cite{DharPhysRevLett2001} and the coupling between the system and the heat bath~\cite{Ariel_Amir}, have been shown to affect heat transport. Particularly, theoretical studies for one dimensional isotopically (mass) disordered harmonic chains show that $J \propto L^{-\frac{1}{2}}$ with free boundary conditions~\cite{Rubin_Greer} while $J \propto L^{-\frac{3}{2}}$ with fixed boundary condition~\cite{Casher_Lebowitz}, implying that $\beta$ can be $\frac{1}{2}$ or $\frac{3}{2}$. For this particular model, it was also shown that normal scaling (i.e. $\beta=1$) can be observed only under specific choices of the thermal bath~\cite{DharPhysRevLett2001}. It was also argued, under free boundary conditions, that one-dimensional harmonic chains with spatially {\it{correlated}} disorder may exhibit normal heat conduction asymptotically~\cite{Herrera_Gonz_lez_2015}.

Can one have normal heat transport in one dimensional disordered (uncorrelated) harmonic chains even within free boundary condition? A recent theoretical study~\cite{Ariel_Amir} predicts that for a {\it{weakly}} coupled disordered harmonic chain one may observe normal heat transport in the presence of {\it{strong disorder}}, when disorder is characterized by a heavy-tailed distribution.  While it is important to verify this theoretical prediction, it is equally interesting to ask: How does thermal conductance scale with $L$ if the coupling between the system and the heat bath is not weak? For a given coupling, how does $\beta$ depend on the nature of the disorder? In this Letter, we address these questions by studying, analytically as well as numerically, the scaling of thermal conductance in one dimensional disordered harmonic chains for different types of disorder and coupling between the system and the heat bath. 

Heat conduction by phonons is similar to electrical conduction, but with a crucial difference: the presence of a localization threshold at zero frequency. This leads to a diverging localization length, $\xi(\omega)$, for $\omega\to 0$~\cite{PhysRevB.28.4106, Ishii} and has strong consequences on the scaling of thermal conductance. Specifically, for a given $L$ and disorder strength, one can define a cut-off frequency $\omega_L$, for which $\xi(\omega_L)=L$. All phonons with $\omega \leq \omega_L$ are effectively delocalized, i.e.~$\xi (\omega) \geq L$, and contribute to the heat transport. 

\begin{figure}[t]
	\includegraphics[width=\columnwidth]{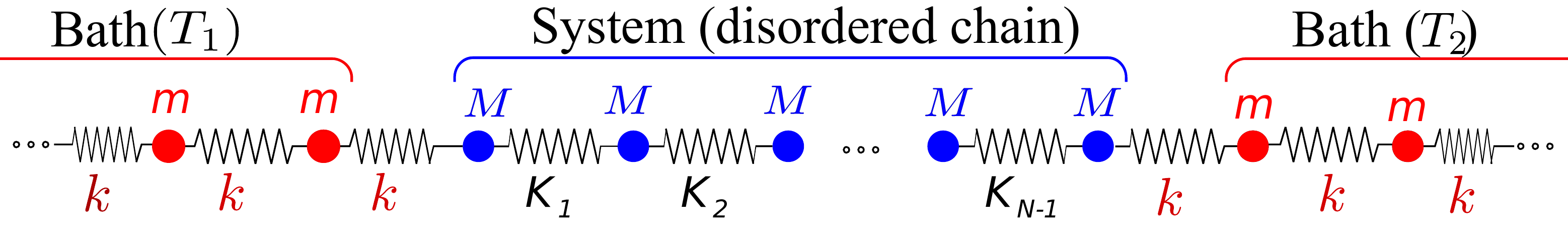}
	\caption{Schematic illustration of one dimensional disordered harmonic chain. Full and empty circles correspond to the bath and system particles, respectively. All particles and springs in the heat baths are identical, with mass $m$ and spring constant $k$. Particles in the system all have mass $M$, and particles $i$ and $(i+1)$ are connected by a springs with stiffness $K_i$. The masses at the two ends of the disordered chain, $i=1$ and $i=N$, are connected to heat baths through a spring of strength $k$.}
	\label{Fig:sketch}
\end{figure}

{\it Model \& background.- }We consider a one dimensional disordered system consisting of $N$ particles, each of mass $M$, connected by harmonic springs with spring constants ${K_i} ( i=1,2,\cdots N-1)$, chosen randomly from a given distribution, cf.~Fig.~\ref{Fig:sketch}. $K_i$ is the spring constant of the spring connecting particles $i$ and $i+1$ in the disordered chain. The two ends $(i=1$ and $i=N)$ of the chain of length $L$ are coupled to two heat baths at temperatures $T_1$ (left bath) and $T_2 (< T_1$; right bath), respectively. Here, $L=(N-1)r_0$ with $r_0$ being the average interparticle distance. Heat baths are modeled as ordered harmonic chains consisting of an infinite number of equal masses $(m)$, and connected by identical springs $(k)$. The system is coupled to two heat baths via two springs each having spring constant $k$. If $k$ is much smaller (larger) compared to the typical spring constant in the disordered chain, we refer to the system as weakly (strongly) coupled to the reservoir. Note that our setup corresponds to the case of `free boundary condition' considered in the literature~\cite{Rubin_Greer,DharPhysRevLett2001}. Below we work in units where the mass $M$ of the system's particles, the natural frequency of the bath $\omega_0=\sqrt{k/m}$, and Boltzmann's constant are all set to unity. We express the stiffness of the springs in units of $M \omega_0^2=1$.

In the current study, we consider two models of disorder: (1) Uniform distribution: $K_i=(1+R_i)$ where $R_i$ follows a uniform distribution of width $W$, i.e., $R_i \in \left[-W/2,W/2 \right]$. Large values of $W$ correspond to stronger disorder and $W=2$ is the strongest possible disorder strength. (2) Power-law distribution: $K_i$  follow a power-law probability distribution, $P(K) \propto K^{\epsilon-1}$, where $0<K\le 1$ and disorder strength is quantified by the dimensionless parameter $\epsilon (>0)$~\cite{FN2}. This situation arises naturally if $K$ decays exponentially with interparticle separation which follows a Poisson process~\cite{Ariel_PRX,Ariel_PRL}. Small $\epsilon(< 1)$ corresponds to strong disorder.

\begin{figure}[t]
	\includegraphics[width=8.5cm,keepaspectratio]{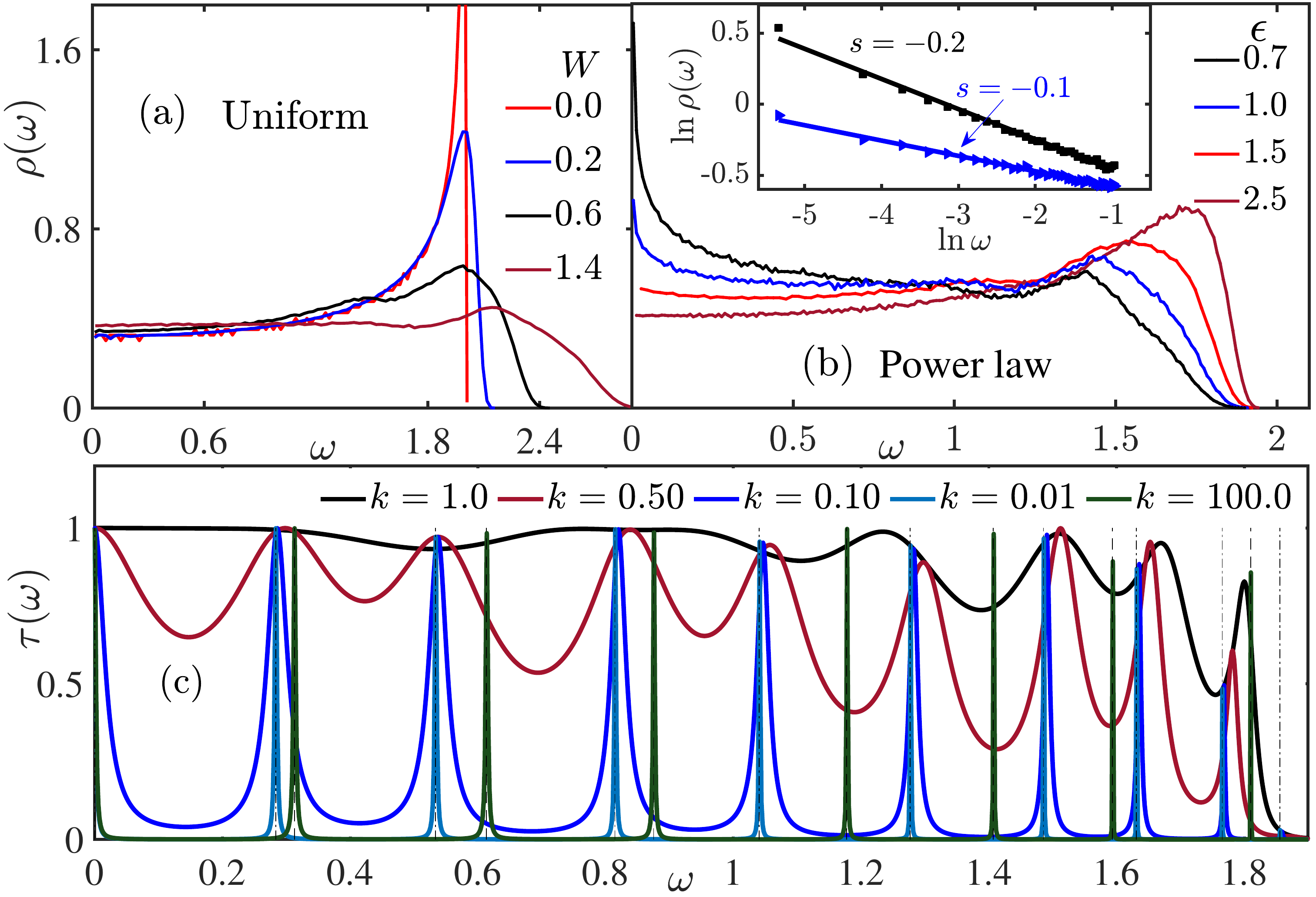}
	\caption{(Color online) 
		Density of states, $\rho(\omega)$, as a function of phonon frequency $\omega$ for different disorder strengths for a one dimensional harmonic chain with (a) uniform disorder and (b) power-law disorder and $N=2000$. (a) For $W=0$, $\rho(\omega)$ diverges when $\omega \to 2$ (a van-Hove singularity) and vanishes for $\omega>2$. Presence of disorder $(W>0)$ smears out the divergence at $\omega=2$ and gives rise to finite $\rho(\omega)$ for $\omega>2$. For $\omega \to 0, \rho(\omega)$ approaches a constant for all disorder strengths. (b) For power-law disorder, $\rho(\omega)$ diverges at $\omega \to 0$ in the strong  disorder regime and approaches a constant in the intermediate and weak regimes (inset). (c) The transmission coefficient $\tau(\omega)$ for various coupling strengths ($k$) and a given realization of disorder. For this example, we consider a chain with uniform disorder ($W=0.5$) and $N=11$. For $k=1.0$ (red line), transmission is roughly constant up to a cutoff frequency $\omega_L$, For $k\ll1$ or $k\gg1, \tau$ also vanishes for $\omega>\omega_L$, but is sharply peaked around the eigenfrequencies of the disordered chain for $\omega<\omega_L$.
	}
	\label{Fig:Fig1}
\end{figure}

\begin{figure}[t!]
\includegraphics[width=8.5cm,keepaspectratio]{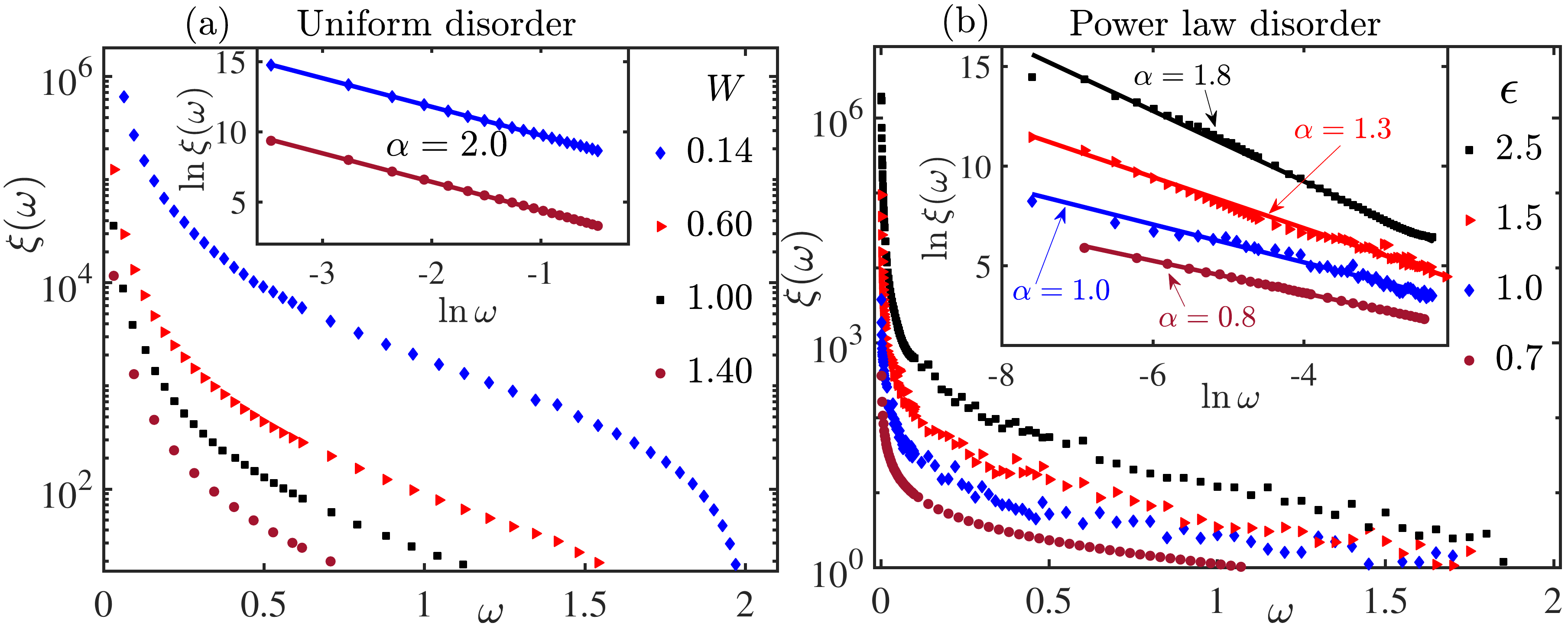}
\caption{(Color online) 
The dependence of localization length, $\xi(\omega)$, on the frequency $\omega$ of the phonons for different strengths of disorder (y-axis in log scale) for (a) uniform disorder and (b) power-law disorder. For small frequencies $(\omega \to 0), \xi(\omega)$ diverges as $\omega^{-\alpha}$ where $\alpha$ depends on disorder strength. (a) For uniform disorder, $\alpha =2$ in weak as well as strong disorder regimes. (b) For power-law disorder, $\alpha \approx 2$ for weak disorder $(\epsilon >2)$ while for intermediate disorder $(1<\epsilon \le 2), \alpha \approx \epsilon$ and in the strong disorder regime $(\epsilon \le 1), \alpha \approx \frac{2\epsilon}{1+\epsilon}$ (inset).
}
\label{Fig:Fig2}
\end{figure}

As noted above, transport is mediated by effectively delocalized low-frequency phonons. Thus, it will be crucial to understand the scaling behavior of the localization length, $\xi(\omega)$ and density of states (DOS), $\rho(\omega)$, in the limit $\omega\to 0$. Here, we briefly summarize earlier theoretical predictions related to these scaling behaviors. With uniform disorder, $\rho(\omega)$ approaches a constant (equivalent to the Debye scaling for 1D ordered systems~\cite{Ashcroft}) while $\xi(\omega) \propto \omega^{-2}$~\cite{RevModPhys.53.175, PhysRevLett.49.337, Ariel_PRL, PhysRevB.28.4106,Ishii}. These results were predicted for weak disorder, but we will show that these hold for any value of $W$. With power law disorder, we define three qualitatively different regimes: in the weak disorder regime $(\epsilon>2)$, $\rho$ and $\xi$ have the same scaling behavior as that of the uniform disorder;  In the intermediate disorder regime $(1<\epsilon \le 2)$, where the variance of the compressibility diverges (but its mean does not), the localization length has a non-trivial scaling $\xi(\omega)\propto \omega^{-\epsilon}$, but the DOS still exhibits Debye scaling, and in the strong disorder regime $(\epsilon \le 1)$, the mean of the system's compressibility diverges, and both $\rho$ and $\xi$ feature anomalous scaling~\cite{PhysRevLett.49.337}: $\rho(\omega) \propto \omega^{\frac{\epsilon-1}{\epsilon+1}}$, $\xi(\omega)\propto\omega^{-\frac{2\epsilon}{\epsilon+1}}$. These results, which are crucial for the discussion below, are summarized in Table~\ref{T:results} and numerically demonstrated in Fig.~\ref{Fig:Fig1}(a-b) and Fig.~\ref{Fig:Fig2}.

{\it Analytical results -} To study the heat transport we follow the Landauer scattering approach in which propagation of a phonon of given frequency $\omega$ through the disordered chain is characterized by a transmission coefficient, $\tau(\omega)$. For a one dimensional system the thermal conductance $G(L,T)$ is~\cite{PhysRevLett.81.232}:
\begin{eqnarray}
G(L,T)\approx \int_{0}^{\infty}\frac{\mathrm{d}\omega}{(2\pi)^2} h \omega \frac{\partial f_{T}(\omega)}{\partial T} \tau(\omega)\ .
\label{Eq:GT1}
\end{eqnarray}
where $f_T$ is the Bose-Einstein distribution function and we also assume $\Delta T=T_1-T_2 \ll T=\frac{T_1+T_2}{2}$.

Note that the system features two competing frequency scales: the disorder-related $\omega_L$ and the thermal frequency $\omega_T=T/\hbar$. Phonons with $\omega> \omega_L$ do not contribute to conductance because they are localized. Phonons with $\omega> \omega_T$ do not contribute because they are not populated. Therefore, the integral in Eq.~\eqref{Eq:GT1} is better represented in terms of the non-dimensional frequency $x=\omega/\omega_T$, 
\begin{align}
G(L,T)=\frac{3g_{qm}}{\pi^2} \int_{0}^{\infty} \mathrm{d}x \frac{x^2\mathrm{e}^x }{\left( \mathrm{e}^x-1 \right)^2 }\tau(x\, \omega_T)\ .
\label{Eq:GT3}
\end{align}
Here $g_{qm}=\pi^2 T/(3h)$ is the quantum of thermal conductance~\cite{PhysRevLett.81.232, Pendry1} which is the universal value of $G(L,T)$ in the limit $T\to 0$. To see this, note that $\tau(\omega)\to 1$ for $\omega\to 0$ due to the existence of a Goldstone mode, related to the translational invariance of the system. Using this fact, it is straightforward to show that for a given $L$ Eq.~\eqref{Eq:GT3} yields $G(L,T)\to g_{qm}$ for very small $T~(\omega_T \ll \omega_L)$, regardless of any other property of $\tau(\omega)$. $g_{qm}$ is thus the natural unit of conductance for our system and below we express all results in these units by defining $G_{qm}(L,T) = G(L,T)/g_{qm}$.  

Of course, the limit of $L\to\infty$ and finite $T$ is of more interest, but an exact evaluation of the integral in the general case is not feasible. Nonetheless, much insight can still be gained in some interesting cases.

We first consider the situation where the stiffness of the coupling spring $k$ is comparable to that of the chain, i.e.~$k\approx 1$~\cite{FN2_2}. In this case there is relatively little impedance mismatch between the chain and the bath, implying less reflectance of the incident phonons from the bath-system boundary. In this situation, phonons get transmitted even when their frequency is not close to an eigenfrequency of the chain. Therefore, from a scaling perspective we can approximate that  $\tau(\omega)=1$ for all phonons with $\omega \leq \omega_L$ and zero otherwise. A numerical calculation of $\tau(\omega)$, shown in Fig.~\ref{Fig:Fig1}c, demonstrates that this approximation is crude but reasonable. As shown below, it quantitatively captures the scaling behavior. 

With this approximation, Eq.~\eqref{Eq:GT1} depends only on the dimensionless combination $\omega_L/\omega_T$ (which is the upper integration limit) implying that thermal conductance for $k\approx1$ should follow a universal curve, independent of temperature and disorder, when expressed in terms of $\omega_L/\omega_T$. In fact the integral can be carried out in closed form, and for large $L$ (or large $T$ .i.e., $\omega_T \gg \omega_L)$ it reads
\begin{align}
G_{qm}(L,T) \approx \frac{3}{\pi^2} \left(\frac{\omega_L}{\omega_T}\right) +\mathcal{O}\left(\frac{\omega_L}{\omega_T}\right)^2\ .
\label{Eq:GT5}
\end{align}
In order to get the explicit dependence on system size in this limit, we use the known scaling $\xi(\omega)\propto  \omega ^{-\alpha}$, see~\cite{Ariel_Amir} and Fig.~\ref{Fig:Fig2}. Straightforward manipulation shows that this implies $G_{qm} \propto L^{-\frac{1}{\alpha}}T^{-1}$, that is,  $\beta=\alpha^{-1}$ in this limit of small impedance mismatch and large $L$. Also, for a given $L$, $G(T) \propto T$ for small $T$ and  $G(T) \sim$ constant (saturates) for high $T$~\cite{supply}.

\newcommand{\bigcell}[1]{\multirow{3}{*}{\makecell{#1}}}
{\renewcommand{\arraystretch}{2}
\begin{table}[]
	\begin{tabular}{p{2cm}cccc}
   \toprule
	Disorder 	& Localization 			& DOS 		&  \makecell{Impedance \\ Mismatch} 				& Conductance 			\\ \hline\hline
	\makecell[l]{Uniform} 		&$\xi\sim \omega^{-2}$ 	& $\rho\sim 1$	& Any &$G\sim L^{-\frac{1}{2}}$\\ \hline
	\makecell[l]{Power law\\$\epsilon\ge2$ (weak)} 	 &$\xi\sim \omega^{-2}$ 		 & $\rho\sim 1$ & Any &$G\sim L^{-\frac{1}{2}}$		\\ \hline
	\makecell[l]{Power law,\\$1\le\epsilon \le 2$}&$\xi\sim \omega^{-\epsilon}$& $\rho\sim 1$& Any &$G\sim L^{-\frac{1}{\epsilon}}$	\\ \hline
	\multirow{2}{*}{\makecell[l]{Power law\\$\epsilon \le 1$ (strong)}}
								&\multirow{2}{*}{$\xi\sim \omega^{-\frac{2\epsilon}{\epsilon+1}}$}&\multirow{2}{*}{ $\rho\sim \omega^{\frac{\epsilon-1}{\epsilon+1}}$}& Low &$G\sim L^{-\frac{1+\epsilon}{2\epsilon}}$\\ 
												& &&  High&$G\sim L^{-1}\vphantom{L^{\frac{1}{2}}}$\\\hline
	\end{tabular}
 \caption{ Summary of the scaling behavior for thermal conductance, $G$ (from this work), density of states (DOS), $\rho$ and localization length, $\xi$ (from ref~\cite{PhysRevLett.49.337}) under different impedance mismatch (coupling $k)$ and disorder strengths.}
\label{T:results}
\end{table}
}
 
For uniform disorder, theory predicts $\alpha=2$ implying $G(L) \propto  L^{-1/2}$, in accord with previously reported results for mass-disordered chain under free boundary condition~\cite{Rubin_Greer, DharPhysRevLett2001}. For power-law disorder, as $\alpha$ depends on disorder strength $\epsilon$, $\beta$ also depends on $\epsilon$ with $G\propto L^{-1/2}, L^{-1/\epsilon}$ and $L^{-\frac{1+\epsilon}{2\epsilon}}$ in the weak, intermediate and strong disorder regimes, respectively. These results are summarized in Table.~\ref{T:results}.

This concludes the case of $k\approx 1$, where transmission is approximately constant for all $\omega$ below a certain cutoff. How does the picture change in the case of strong impedance mismatch, $k\gg1$ or $k\ll1$? In this case transmission is negligible for almost all frequencies, except those which are close to an eigenfrequency of the disordered chain. In previous work~\cite{Ariel_Amir}, it was shown that in the weak coupling limit, $k \ll 1$, $\tau$ has a structure of non-overlapping Lorentzians for phonons with $\omega<\omega_L$, cf.~Fig.~\ref{Fig:Fig1}c. Each Lorentzian is centered around an eigenfrequency of the disordered chain and the area of each Lorentzian, i.e. it's integrated contribution to the thermal conductance, was shown to be $\omega$-independent for the delocalized modes~\cite{Ariel_Amir}. Calculating the integral in general for any $\omega_L/\omega_T$ is difficult, but if we are only interested in the scaling behavior for large $L$, the integral essentially counts the number of eigenmodes of the disordered chain within the frequency range $0<\omega \le \omega_L$~\cite{FN3}:
\begin{align}
G(L) \approx \Sigma \int_0^{\omega_L} \mathrm{d}\omega~\rho(\omega) \ ,
\label{Eq.clpse_smlK}
\end{align}
where $\Sigma$ is the area of each Lorentzian. Considering $\rho(\omega)=D\omega^s$ (see~\cite{Ariel_Amir} and Fig.~\ref{Fig:Fig1}a-b), where $D$ depends on disorder, for large $L$ we get
\begin{align}
{G} &\propto  \omega_L^{s+1} \propto  L^{-\frac{s+1}{\alpha}}.
\label{Eq.clpse_smlK2}
\end{align}

Thus, in the weak coupling regime $G_{qm} \propto L^{-\frac{s+1}{\alpha}}T^{-1}$ for a fixed $T$. For uniform disorder, as well as power-law disorder with $\epsilon>1$ (i.e.~the weak and intermediate regimes), we have $s=0$ and thus the scaling of thermal conductance with $L$ remains the same as in case of an impedance-matched bath $k\approx 1$. Interestingly, for strong disorder, $\epsilon<1$, we have $s=\frac{\epsilon-1}{\epsilon+1}$ and $\alpha=\frac{2\epsilon}{1+\epsilon}$, which together cancel out exactly to yield normal Fourier-like heat conduction $\beta=1$. Also, note that, when expressed in terms of $\omega^{s+1}_L/\omega_T$, thermal conductance should follow a universal curve in the weak coupling regime and the large $L$ limit.

Lastly, we deal with the case of very large $k$, i.e.~the strong coupling regime. A careful analysis, presented fully in the supplementary material~\cite{supply}, shows that this limit is equivalent to a system with the first and last particles excluded, i.e., effectively a system of $(N-2)$ particles. Therefore, like in the case of weak coupling, $\tau(\omega)$ is composed of non-overlapping peaks with an $\omega$-independent area, cf.~Fig.~\ref{Fig:Fig1}. Since the density of states and the localization length are independent of the coupling $k$, all our predictions for $k\ll 1$ hold also for $k\gg 1$. That is, the same scaling exponents emerge in the case of strong impedance mismatch, regardless of whether $k$ is very small or very large. All our theoretical predictions for different disorder types and coupling strengths are summarized in Table.~\ref{T:results}.

\begin{figure}[t!]
\includegraphics[width=8.5cm,keepaspectratio]{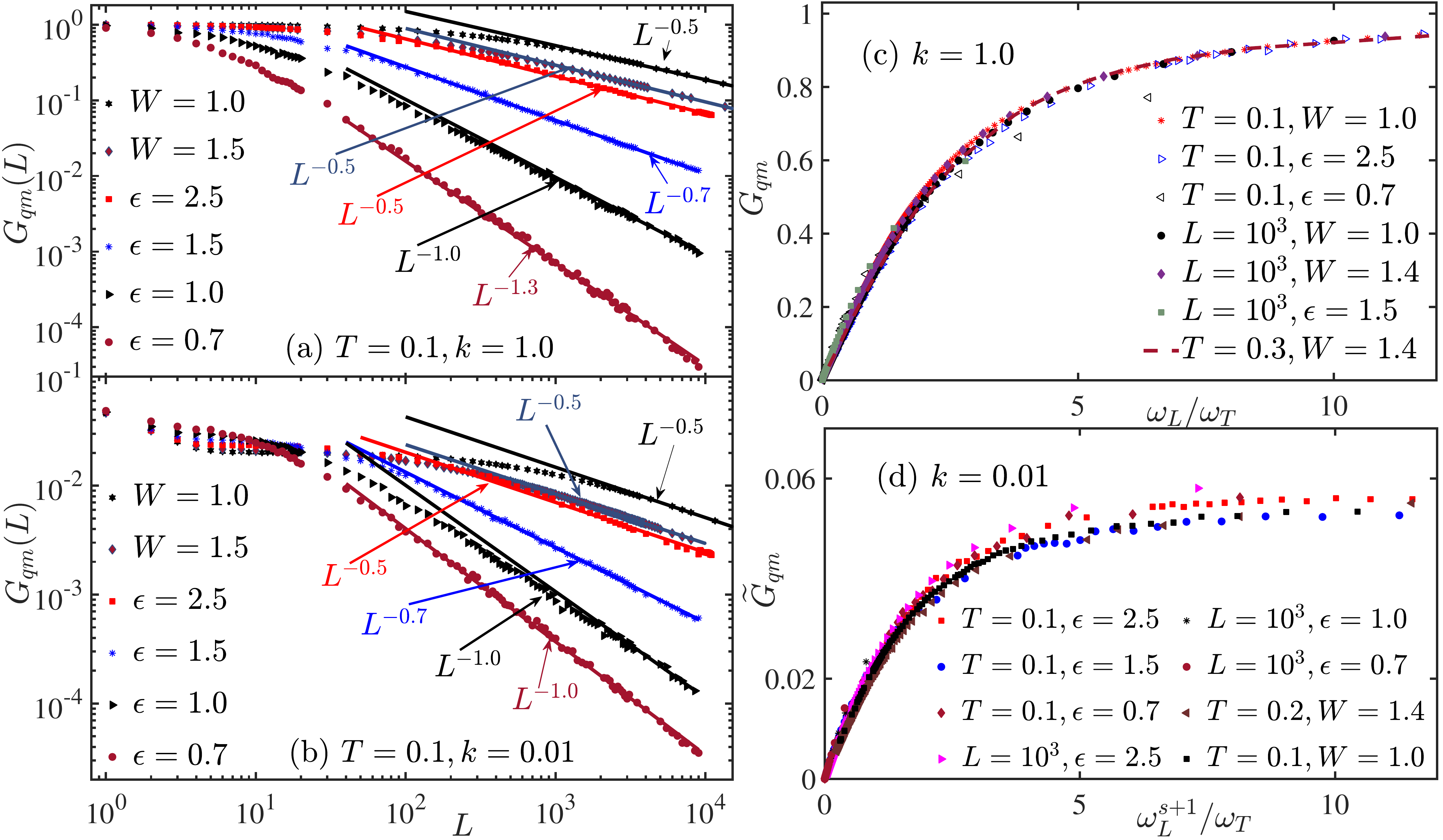}
\caption{(Color online) 
Dependence of the thermal conductance $G_{qm}(L)$ on the length $L$ of the disordered chain for uniform (disorder strength $W$) and power-law (disorder strength $\epsilon$) disorder in a log-log plot with (a) $k=1.0$ and (b) $k=0.01$. The mean temperature is fixed at $T=0.10$. For $k=1.0$, $G_{qm}(L) \approx 1$ for very small $L$, while $G_{qm}(L) \propto L^{-\beta}$ for  large $L$. $\beta \ne 1$ (anomalous scaling) for $k=1.0$ except for $\epsilon=1$. For $k=0.01, \beta=1.0$ in the strong disorder regime $(\epsilon \le 1.0)$ while in all other cases $G_{qm}(L)$ scales anomalously.
Unified description of thermal conductance for one dimensional disordered harmonic chains with different models of disorder in terms of the variable (c) $\omega_L/\omega_T$ (for $k=1.0$) and (d) $\omega_L^{s+1}/\omega_T$ (for $k=0.01$).
Here, $\tilde{G}_{qm}=\frac{(s+1)G_{qm}}{D}$ (see text for details).
}
\label{Fig:Fig3}
\end{figure}

{\it Numerical results.-} We test our theoretical predictions by numerically computing thermal conductance and other properties, such as density of states and localization lengths, for all cases considered above. The density of states, $\rho(\omega)$, for different disorder strengths is shown in Fig.~\ref{Fig:Fig1}(a-b) for the uniform and power-law disorder, respectively, with $N=2000$~\cite{supply}. For uniform disorder, $\rho(\omega)$ approaches a constant as $\omega \rightarrow 0$, that is, $s=0$. For power law disorder, $\rho(\omega)$ diverges with an exponent consistent with the theoretical prediction, $s=\frac{\epsilon -1}{\epsilon +1}$. For weak and intermediate disorder theory predicts $s=0$ but a weak divergence is observed for $\epsilon=1$, the origin of which in not clear to us at present.

To calculate $G$, we compute $\tau(\omega)$ directly for different $k$ and disorder types using a transfer matrix method~\cite{supply}. For a single realization of the disorder, the dependence of $\tau(\omega)$ on $k$ is shown in Fig.~\ref{Fig:Fig1}(c) (with $W=0.50$ and $N=11$). 

For a given disorder and a fixed $\omega$, we find that $\tau(\omega)$ decays exponentially with $L$~\cite{supply}. This defines a length scale which we interpret as the localization length $\xi$, i.e., $\tau(\omega, L) \propto \exp[-L/\xi(\omega)]$~\cite{supply}. We find that $\xi$ diverges like $\xi(\omega) \propto \omega^{-\alpha}$, consistent with  theoretical predictions (see Fig.~\ref{Fig:Fig2}).

Finally, we compute $G$ for different $L$, $T$, disorder types and coupling strengths, using Eq.~\eqref{Eq:GT1}. This is presented in Fig.~\ref{Fig:Fig3}(a)  (for $k=1.0$) and \ref{Fig:Fig3}(b) (for $k=0.01$), which shows that the numerical results for all cases agree with the theoretical predictions (for $k=100$, see supplemental material~\cite{supply}). Heat transport is anomalous $(\beta \neq 1)$ for all cases except for the power-law disorder in the weak/strong coupling regime and also when $\epsilon=1$ (Fig.\ref{Fig:Fig3}(b)). In addition, panels (c) and (d) demonstrate that when expressed in terms of $\omega_L/\omega_T$ (for $k =1$) and $\omega_L^{s+1}/\omega_T$ (for $k=0.01)$, respectively, all data collapse on a single curve, following Eq.~\eqref{Eq:GT5} and Eq.~\eqref{Eq.clpse_smlK2}.

{\it{Conclusions and discussion.- }}In this Letter we studied analytically the thermal conductance of one-dimensional disordered harmonic chains and corroborated the theoretical predictions with extensive numerical simulations. We found a non-trivial scaling behavior of the conductance, which depends both on the nature of disorder and the coupling between the system and the heat baths. This dependence is mediated by the scaling of the localization length and density of states. In addition, we identified the dimensionless scaling parameter with which one has a unified description for all temperatures and systems sizes. 

Specifically, we found that the conductance features anomalous scaling with $L$ for uniform disorder and for weak and intermediate power-law disorder (i.e. with a well-defined mean). Interestingly, for strong power-law disorder and strong impedance mismatch, $k\ll1$ or $k\gg1$, normal scaling $G\propto L^{-1}$ is observed. For strong disorder and low impedance mismatch, i.e.~$k\approx 1$, the scaling exponent $\beta$ can be greater than unity. While most of the previous works on anomalous heat transport focused on the classical regime, our study encompasses both, quantum and classical regimes.
It would be interesting to study in the future the fluctuations in the thermal conductance, and in particular ask if the universal fluctuations observed in electronic systems have a counterpart here.

{\it{Acknowledgements.- }}We thank David Mross for valuable discussions. BA acknowledges the computational facilities at the Weizmann Institute of Science and Harvard University. Financial support from the Weizmann Institute of Science is gratefully acknowledged. YBS was supported by the JSMF post-doctoral fellowship for the study of complex systems.

\bibliography{References}

\begin{thebibliography}{40}
\bibitem[{FN2({\natexlab{a}})}]{FN2}
\bibinfo{note}{Note that $K$ is bounded from above and the power law tail is at
  low values of $K$.}
\bibitem[{FN2({\natexlab{b}})}]{FN2_2}
\bibinfo{note}{In our model we always have $M=1$ for the masses comprising the
  disordered system. For a fixed disorder strength, when we vary $k$ to study
  the effect of coupling strength, we also vary $m$ with it to maintain the
  natural frequency of the ordered chain as $\omega_0=\sqrt{\frac{k}{m}}=1$.}
\bibitem[{FN3()}]{FN3}
\bibinfo{note}{For large $L$, only low frequency modes $(\omega \to 0)$
  contribute to the heat transport. Thus, in this limit, the mean temperature
  of the baths is effectively infinte $(T \gg \hbar\omega)$ and becomes
  irrelevant~\cite{Ariel_Amir}}.
\end{thebibliography}
\bibliographystyle{apsrev}


\clearpage
\onecolumngrid
\renewcommand{\theequation}{S.\arabic{equation}}
\setcounter{equation}{0}

\renewcommand{\thefigure}{S.\arabic{figure}}
\setcounter{figure}{0}

\section*{Supplemental material for `Thermal conductance for one dimensional disordered harmonic chains'}

\section{Analytical prediction of the transmission coefficient in the regime of strong impedance mismatch}
\label{Sec.General}
The model of a one dimensional disordered chain with $N$ particles coupled to two heat baths at the two ends is shown in Fig.~1 of the main article. $u_n$ denotes the displacement of the $n$-th particle in the disordered chain from its equilibrium position. The equation of motion of the $n$-th particle is
\begin{align}
M\frac{\mathrm{d}^2u_n}{\mathrm{d}t^2} = K_{n}(u_{n+1}-u_n) - K_{n-1}(u_{n}-u_{n-1})
\label{Eq.EOM1}
\end{align}
Considering $u_n=x_n \exp[-i\omega t]$ (i.e. looking at normal mode solution) in Eq.~\eqref{Eq.EOM1}, we get
\begin{align}
-M \omega^2 x_n = -K_{n-1} (x_n - x_{n-1}) + K_n (x_{n+1}-x_n)
\label{Eq:EOM2}
\end{align}
Note that $K_0=k,$ and $K_N=k$. We can write the above set of equations in a compact form using the following matrix notation:
\begin{align}
M \omega^2 
\begin{bmatrix} 
1       &   0      & \cdots & 0       \\
0       &   1      & \cdots & 0       \\
\vdots  &   \vdots & \ddots & \vdots  \\
0       &   0      & \cdots & 1       \\
\end{bmatrix}
\begin{bmatrix} 
x_1     \\
x_2     \\
\vdots  \\
x_N 
\end{bmatrix} &+
\begin{bmatrix} 
-k -K_{1}   &  K_{1}         &  0             &  0     		    & \ldots   & 0                  &  0           \\
K_{1}       & -K_{1} -K_{2}  & K_{2}          &  0                  & \ldots   & 0                  &  0           \\
0           &  K_{2}         & -K_{2} -K_{3}  & K_{3}                & \ldots   & 0                  &  0           \\
0           &  0             &  K_{3}         & -K_{3} -K_{4}    & \ldots   & 0                  &  0           \\
\vdots      & \vdots         & \cdots         & \cdots &  \ddots    & \vdots             & \vdots       \\
0           & 0              & 0              & \cdots &  \vdots    & -K_{N-2} -K_{N-1}  &  K_{N-1}     \\
0           & 0              & 0              & \cdots &  \vdots    &  K_{N-1}           & -K_{N-1} -k
\end{bmatrix}
\begin{bmatrix} 
x_1     \\
x_2     \\
x_3     \\
x_4     \\
\vdots  \\
x_{N-1} \\
x_N 
\end{bmatrix}
\nonumber\\
&=
\begin{bmatrix}
-k x_{0}   \\
0              \\
\vdots         \\
0              \\
-k x_{N+1} 
\end{bmatrix}  
\equiv
\begin{bmatrix}
b_1     \\
b_2     \\
\vdots  \\
b_{N-1} \\
b_{N} 
\end{bmatrix} \nonumber\\
\text{or,~~}  \left( M \omega^2 I + A \right) |X\rangle &= |B \rangle \nonumber \\
\text{or,~~}  |X\rangle &= \left( M \omega^2 I + A \right)^{-1} |B \rangle ,
\label{Eq:EOMM}
\end{align}
where, $I$ is a unit matrix of order $N \times N$, and $A$ is a tri-diagonal matrix of order $N \times N$ with $A_{i,i}=-K_{i-1} -K_{i}, A_{i,i+1}=K_{i}$ and $A_{i,i-1}=K_{i-1}$. Also, $|X\rangle$ and $|B\rangle$ are column vectors and all components of $|B\rangle$ are zero except for the first and last elements; $b_1=-k x_{0}$ and $b_N=-k x_{N+1}$. We are using the quantum-mechanical bra-ket notation for convenience.

As the normalized eigenvectors $\{|V_i \rangle\}$, corresponding to the eigenvalues $\{ \lambda_i \} (i=1,2,\cdots,N)$ of $A$ form a complete basis, we can expand the right hand side of Eq.~\eqref{Eq:EOMM} in terms of $\{|V_i \rangle\}$ and write the Eq.~\eqref{Eq:EOMM} as
\begin{align}
|X\rangle = \sum_{i=1}^{N}\frac{1}{M\omega^2+\lambda_i}|V_i \rangle \langle V_i | B \rangle 
\label{Eq:EOMM2}
\end{align}
Whenever there is strong impedance mismatch between the bath and the system (which occurs in case of weak and strong coupling), transmission is appreciable only for frequencies close to the eigenfrequencies of $A$. In this case the sum in Eq.~\eqref{Eq:EOMM2} is dominated by a small number of summands. 

An eigenmode can have a non negligible contribution to the sum of Eq.~\eqref{Eq:EOMM2} in one of two cases: either
\begin{enumerate}[(a)]
	\item it is close to resonance, i.e.~the denominator $M\omega^2+\lambda_i$ is small, or 
	\item the projection of the eigenmode on the chain ends, $\left\langle V_i | B \right\rangle$ is large. 
\end{enumerate} 

As was shown in ref.~\cite{Ariel_Amir}, in the case of very weak coupling ($k\ll1$) only case (a) occurs and near an eigenfrequency of the chain Eq.~\eqref{Eq:EOMM2} is well approximated by 
\begin{align}
|X\rangle \approx \frac{1}{\delta \lambda_i}|V_i \rangle \langle V_i | B \rangle 
\label{Eq.EOM3}
\end{align}
where, $\delta \lambda_i = M (\omega^2 - \omega_i^2)$ as $\lambda_i = -M\omega_i^2$. Using Eq.~\eqref{Eq.EOM3}, one can derive an expression for transmission coefficient as described in ref.~\cite{Ariel_Amir}, which is valid for weak coupling.

The case of very large coupling $(k\gg1)$ requires more care as both options (a) and (b) occur. In addition to resonant modes, the strong springs in the ends of the chains give rise to two special modes, each localized in one of the ends of the chain. These modes' contribution to the sum of Eq.~\eqref{Eq:EOMM2} is not negligible for any frequency (with respect to the contributions from other modes).

To see this, let us order the eigenvalues of $A$ in ascending order of their magnitude: $\lambda_1 < \lambda_2 < \cdots < \lambda_i < \cdots <\lambda_{N-1} <\lambda_{N}$. In Sec.~\ref{Sec.DegPrertb} we show, within a rigorous degenerate perturbation theory, that for $k \gg 1$ these two special modes are $|V_{N-1}\rangle \approx [1~0~0 ~\cdots ~0]^T$ and  $|V_{N}\rangle \approx [0~ 0~ 0~ \cdots~ 1]^T$, and their associated eigenvalues are $\lambda_{N-1} \approx \lambda_{N} \approx -k$. All other eigenvalues are much smaller.

Thus, in the strong coupling limit, instead of Eq.~\eqref{Eq.EOM3}, we have
\begin{align}
|X\rangle \approx \frac{1}{\delta \lambda_i}|V_i \rangle \langle V_i | B \rangle + \frac{1}{\delta \lambda_{N-1}}|V_{N-1} \rangle \langle V_{N-1} | B \rangle + \frac{1}{\delta \lambda_{N}}|V_{N} \rangle \langle V_{N} | B \rangle.
\label{Eq.EOMN3}
\end{align}

Writing $\omega_i=\bar{\omega}$ and $\delta \lambda_{i} = \delta \lambda$, thus, the displacements $x_1$ and $x_N$ are given by
\begin{align}  
x_1 &\approx \frac{-k x_0 v_1^2   - k x_{N+1}v_1 v_N}{\delta \lambda} - \frac{k}{\delta \lambda_{N-1}} x_0\ , &
x_N &\approx \frac{-k x_0 v_1 v_N - k x_{N+1}v_N^2}{\delta \lambda} - \frac{k}{\delta \lambda_N} x_{N+1}\ ,
\label{Eq:EEq1}
\end{align}
where, $v_j\equiv\braket{j|v}$ is the amplitude of the eigenmode $\bar{\omega}$ at site $j$. Assuming that a phonon of frequency $\omega$ (with unit amplitude) and wave number $q$ coming from the left side of the disordered chain, we have $x_n = \mathrm{e}^{i n\chi} + r \mathrm{e}^{-in\chi} (n<1)$ and $x_{m}=t \mathrm{e}^{i m\chi} (m>N)$, where $r$ and $t$ represent the reflection and transmission amplitudes and $\chi=qa$, $a$ being the lattice constant in the ordered region. Using phonon dispersion relation~\cite{Ashcroft}, we have
\begin{align}
\chi=\cos^{-1} \left(1-\frac{\omega^2}{2\omega_0^2}\right).
\label{Eq:Dsp1}
\end{align}
Note that $x_{N+1}=t$ and $x_0=1+r$. Substituting these expressions for $x_0$ and $x_{N+1}$ in Eq.~\eqref{Eq:EEq1} and using Eq.~\eqref{Eq:EOM2}, we can find the transmission amplitude. In order to simplify the algebra, we assume $\lambda_N = \lambda_{N-1} \approx -k$, so that $\delta \lambda_{N} = M \omega^2 + \lambda_{N} = \delta \lambda_{N-1}$. With this, we get the following expression for  transmission amplitude: 
\begin{align}  
t &= \frac{k^3 v_1v_N \delta \lambda_N ^2 \mathrm{e}^{-i\chi}(\mathrm{e}^{2i\chi}-1)}{\left[ (\mathrm{e}^{i\chi}-2)k \delta\lambda_N + \delta \lambda_N m \omega^2 - k^2 \right]
\left[  (\mathrm{e}^{i\chi}-2)k \delta \lambda_N \delta \lambda -k^2 \left( \delta \lambda + \delta \lambda_N(v_1^2 + v_N^2 \right) + \delta \lambda_N \delta \lambda m \omega^2 \right]}.
\label{Eq:TEq1}
\end{align}
For large $N$, as only low frequency modes contribute to the transmission coefficient $\chi \ll 1$ leading to  
$\chi=qa\approx \omega \sqrt{\frac{m}{k}}$. For very small $\chi$, the transmission coefficient takes the following form
\begin{align}  
\tau = |t|^2 &\approx \frac{4 k^2 (v_1v_N)^2 \chi^2 \delta \lambda_N^4}
{
\left[ \chi^2 \delta\lambda_N^2 + (k + \delta\lambda_N - \delta \lambda_N \chi^2)^2 \right]
\left[ \delta \lambda_N^2 \delta \lambda^2 \chi^2 + \left\{ k \left( \delta \lambda + \delta \lambda_N(v_1^2 + v_N^2) \right) - \delta \lambda_N \delta \lambda ( \chi^2 -1)\right\}^2 \right]} \nonumber \\
&= \frac{4 k^2 (v_1v_N)^2 \chi^2}
{
\left[ \chi^2 + (\frac{k}{\delta\lambda_N} + 1 - \chi^2)^2 \right]
\left[ \delta \lambda^2 \chi^2 + \left\{ k \left( \frac{\delta \lambda}{\delta\lambda_N} + (v_1^2 + v_N^2) \right) - \delta \lambda ( \chi^2 -1)\right\}^2 \right]}.
\label{Eq:TEq2}
\end{align}
As $\omega$ is very small and $\lambda_N \approx -k$, we can write 
\[
\frac{k}{\delta\lambda_N} = \frac{k}{\lambda_N + M \omega^2} \approx \frac{k}{\lambda_N} \approx -1.
\]
Using this, we have
\begin{align}  
\chi^2 + \left(\frac{k}{\delta\lambda_N} + 1 - \chi^2 \right)^2 & \approx \chi^2 \\
\delta \lambda^2 \chi^2 + \left\{ k \left( \frac{\delta \lambda}{\delta\lambda_N} + (v_1^2 + v_N^2) \right) - \delta \lambda ( \chi^2 -1)\right\}^2 &\approx \chi^2 (\delta \lambda - s)^2 + s^2(1-\chi^2),
\label{Eq:TEq3}
\end{align}
where, $s=k(v_1^2 + v_N^2)$. Using above simplifications, finally, we can write
\begin{align}  
\tau &\approx \frac{4 k^2 (v_1v_N)^2}
{\chi^2\left[(\delta \lambda - s)^2 + \frac{s^2(1-\chi^2)}{\chi^2}\right]} \nonumber \\
& \approx \frac{4 k^2 (v_1v_N)^2/\chi^2}
{\left[(\delta \lambda - s)^2 + \frac{s^2}{\chi^2}\right]}.
\label{Eq:TEq4}
\end{align}

\begin{figure}[t!]
	\includegraphics[width=12cm, keepaspectratio]{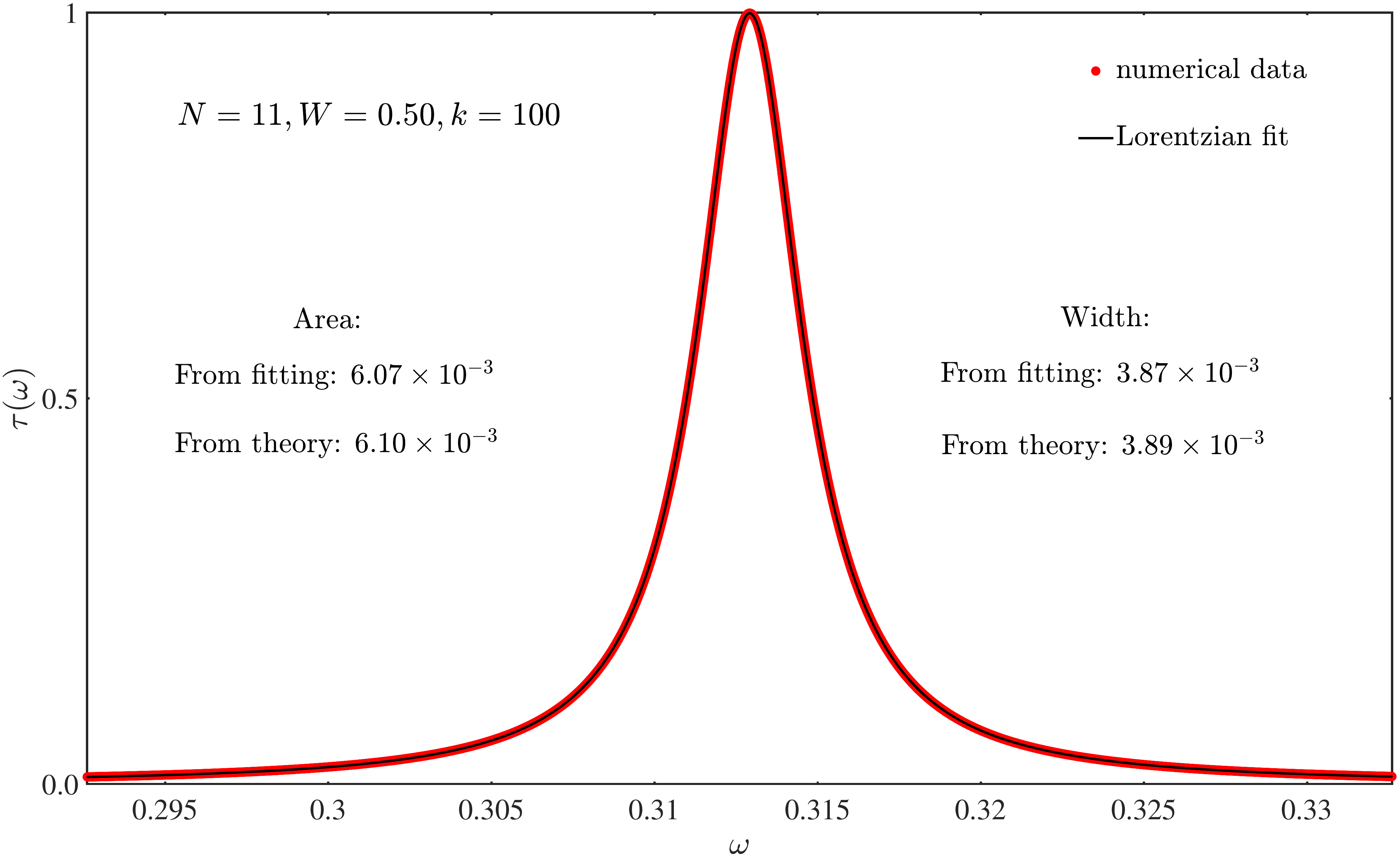}
        \caption{ Transmission coefficient $\tau(\omega)$ for an arbitrarily chosen frequency range for a one dimensional disordered chain with $N=11, W=0.5$ (uniform disorder), and $k=100$, depicting that $\tau(\omega)$ is a Lorentzian in terms of $\omega$. The thick dots represent the numerical data which are same as in Fig.~2c of the main article. The area and width of the Lorentzian agree well with the theoretical predictions (see Eq.~\eqref{Eq.Larea} and Eq.~\eqref{Eq.Lwdth} in this supplemental material).
}
	\label{Fig:Lfit}
\end{figure}
We define a frequency $\widetilde{\omega}$ such that $\delta \lambda - s = M (\omega^2 - \widetilde{\omega}^2)$. As transmission is appreciable for frequencies close to the eigenfrequencies of the chain, we have $\omega \approx  \widetilde{\omega} \approx \bar{\omega}$ and consequently $ \delta \lambda - s \approx 2 M \widetilde{\omega} (\omega - \widetilde{\omega})$. Substituting this into Eq.~\eqref{Eq:TEq4} shows that the transmission coefficient is approximately a Lorentzian in terms of $\omega:$
\begin{align}  
\tau(\omega) & \approx \frac{4 k^2 (v_1v_N)^2/\chi^2}
{\left[(2 M \widetilde{\omega})^2 (\omega - \widetilde{\omega} )^2 + \frac{s^2}{\chi^2}\right]} =\frac{\gamma}{\pi} \frac{\frac{\sigma}{2}}
{(\omega - \widetilde{\omega} )^2 + (\frac{\sigma}{2})^2}.
\label{Eq:TEq5}
\end{align}
where the area $(\gamma)$ and width $(\sigma)$ are given by
\begin{align}
\gamma = \frac{2 \pi (v_1 v_N)^2 k}{M \widetilde{\omega} (v_1^2 + v_N^2)\chi} = \frac{2 \pi (v_1 v_N)^2 k}{M \widetilde{\omega}^2 (v_1^2 + v_N^2)}\sqrt{\frac{k}{m}} \label{Eq.Larea},\\
\sigma = \frac{k (v_1^2 + v_N^2)}{M \widetilde{\omega}\chi} = \frac{k (v_1^2 + v_N^2)}{M \widetilde{\omega}^2}\sqrt{\frac{k}{m}}.
 \label{Eq.Lwdth}
\end{align}   
For large $k$, we compute $v_1$ and $v_N$ using perturbation theory and interestingly,  find (see sec.~\ref{Sec.DegPrertb}) that for the  low-frequency delocalized modes $|v_1|\approx |v_N| \sim \frac{\omega}{k\sqrt{N}}$ and subsequently $\gamma \sim \frac{2 \pi }{\sqrt{km}MN} $, i.e., for the delocalized modes area associated with each Lorentzian is independent of the frequency. Thus, their contribution to the thermal conductance is constant, i.e., independent of the frequency.
 
We also note that the equations for $\gamma$ and $\sigma$ for large $k$ are slightly different from those obtained for small $k (\ll 1)$ where the area $(\gamma_s)$ and the width $(\sigma_s)$ of each Lorentzian are given by (see eq.~15 in ref.~\cite{Ariel_Amir}):
\begin{align}
\gamma_s = \frac{2 \pi (v_1 v_N)^2 k \chi}{M \widetilde{\omega} (v_1^2 + v_N^2)} = \frac{2 \pi (v_1 v_N)^2 }{M(v_1^2 + v_N^2)}\sqrt{km} \label{Eq.sarea},\\
\sigma_s = \frac{k \chi(v_1^2 + v_N^2)}{M \widetilde{\omega}} = \frac{(v_1^2 + v_N^2)}{M }\sqrt{km}.
 \label{Eq.swdth}
\end{align}   
For small $k, \chi$ appears in the numerator of the expressions for the area and the width. In the weak coupling scenario $|v_1|=|v_N|=\frac{1}{\sqrt{N}}$ for the delocalized modes and thus one finds $\gamma_s = \frac{2 \pi \sqrt{km}}{MN}$~\cite{Ariel_Amir}, i.e., the area of each Lorentzian associated with the delocalized modes are independent of the frequency. 

To verify that $\tau(\omega)$ is indeed Lorentzian in $\omega$ with area and width given by Eq.~\eqref{Eq.Larea} and Eq.~\eqref{Eq.Lwdth} for large $k$, we fit the numerical data for $\tau(\omega)$ (see Fig.~2c in the main article) with a Lorentzian given in Eq.~\eqref{Eq:TEq5}. As shown in Fig.~\ref{Fig:Lfit}, we find that the area and the width given by Eq.~\eqref{Eq.Larea} and Eq.~\eqref{Eq.Lwdth}, respectively, agree well with those obtained by fitting the numerical data.

\section{Characteristics of the eigenmodes and the eigenfrequencies of $A$ in the strong coupling limit}
\label{Sec.DegPrertb}

In Sec.~\ref{Sec.General} we used two properties of the eigenmodes in the $k\gg 1$ limit: 
\begin{itemize}
	\item There are two boundary modes (that is, eigenmodes localized at one of the ends of the system) with eigenvalue $\approx k$ that does not scale with $N$. 
	\item For eigenmodes with $\omega\to0$ the vibration amplitude of the first and last particles is proportional to $\omega$. 
\end{itemize}
The former was used to approximate Eq.~\eqref{Eq.EOMN3} from Eq.~\eqref{Eq:EOMM2}. The latter was used to show that the contribution of each delocalized mode to the condactuance is $\omega$-independent in the $\omega\to0$ limit, Eq.~\eqref{Eq.Lwdth}. Here we derive both results within first order degenerate perturbation theory.

To set the ground, we write $A$ as:
\begin{align}
A &= 
\begin{bmatrix} 
-k -K_{1}   &  K_{1}         &  0             &  0     		&  ~~0     & \ldots   & 0                  &  0           \\
K_{1}       & -K_{1} -K_{2}  & K_{2}          &  0              &  ~~0     & \ldots   & 0                  &  0           \\
0           &  K_{2}         & -K_{2} -K_{3}  & K_{3}           &  ~~0     & \ldots   & 0                  &  0           \\
0           &  0             &  K_{3}         & -K_{3} -K_{4}   &  ~~K_{4} & \ldots   & 0                  &  0           \\
\vdots      & \vdots         & \cdots         & \cdots          &  \ddots & ~~\vdots   & \vdots             & \vdots       \\
0           & 0              & 0              & 0               &  0       & \cdots   & -K_{N-2} -K_{N-1}  &  K_{N-1}     \\
0           & 0              & 0              & 0               &  0       & \cdots   &  K_{N-1}           & -K_{N-1} -k
\end{bmatrix} = A^b + A^d, 
\label{Eq.DPT0}
\end{align}
where, 
\begin{align}
\hspace{-2.1cm} A^b&= 
\begin{bmatrix} 
-k        &   0      &  \cdots       	&  \cdots  &     0       		\\
0             &   0      &  0      	&  \cdots  &     0       		\\
0             &   0      &  0      	&  \cdots  &     0       		\\
0             &   0      &  0      	&  \cdots  &     0       		\\
\vdots        & \cdots   &  \cdots       	&  \ddots  &     \vdots  		\\
0             &   0      &  0      	&  \cdots  &     0       		\\
0             &    0     &  0     	&  \cdots  &    -k       	\\
\end{bmatrix},  \nonumber \\
\hspace{-2.1cm} A^d &=
\begin{bmatrix} 
-K_{1}   &  K_{1}         &  0             &  0     		&  0     & \ldots   & 0                  &  0           \\
K_{1}       & -K_{1} -K_{2}  & K_{2}          &  0              &  0     & \ldots   & 0                  &  0           \\
0           &  K_{2}         & -K_{2} -K_{3}  & K_{3}           &  0     & \ldots   & 0                  &  0           \\
0           &  0             &  K_{3}         & -K_{3} -K_{4}   &  K_{4} & \ldots   & 0                  &  0           \\
\vdots      & \vdots         & \vdots         & \ddots          &  \vdots & \vdots   & \vdots             & \vdots       \\
0           & 0              & 0              & 0               &  0       & \cdots   & -K_{N-2} -K_{N-1}  &  K_{N-1}     \\
0           & 0              & 0              & 0               &  0       & \cdots   &  K_{N-1}           & -K_{N-1}
\end{bmatrix}. \nonumber
\end{align}
In the strong coupling limit $(k \gg 1)$ as the typical value of $K_i$ is much smaller than $k$, we consider $A^d$ as a perturbation and compute the first order correction to the eigenvalues and eigenvectors of the unperturbed matrix $A^b$. We see that $A^b$ has two eigenspaces (``degenerate bands''): The first consists of two modes, $\ket{1}=[1~0~0~\cdots~0]^T$ and $\ket{N}=[0~0~0~\cdots~1]^T$, localized at the first and last particle, respectively. Both have eigenvalue $-k$.  The second eigenspace consists of $(N-2)$ modes with $\omega = 0$. This eigenspace consists of all vibrations that keep the first and the last particles fixed.

First, we can already conclude that the two boundary modes with eigenvalue $\approx -k$ indeed exist. To zeroth order their eigenvalue is exactly $-k$ and they are entirely localized at the first and last particles. Any corrections to this picture become smaller as $k$ increases and thus the perturbation becomes smaller.

We turn to analyze the modes in the $(N-2)$-fold degenerate band. Here the zeroth order picture is trivial and we need the first order correction in order to see the structure. Degenerate perturbation theory~\cite{Sakurai} tells us that the relevant eigenmodes of the perturbed system are those which diagonalize the perturbation when projected to the degenerate eigenspace. In our case, we need to find the eigenmodes of the matrix that is obtained by deleting the first and last row and column from $A^d$, namely
\begin{align}
\widetilde{A}& = \begin{bmatrix} 
-K_{1} -K_{2} &  K_{2}       	&  0            &  0       & \cdots     &  0 &  0       		 \\
K_2           & -K_{2} -K_{3}   &  K_3          &  0       & \cdots     &  0 &  0 		 \\
0             &   K_3           & -K_{3} -K_{4} &  K_4     &\cdots      & \vdots & \vdots     	 \\
\vdots        &   \cdots        & \cdots        & \cdots   & \ddots     & \cdots & \cdots	 \\
0             &   0             & \cdots        & \cdots & \cdots   &-K_{N-3}-K_{N-2} &  K_{N-2}            \\
0             &   0             &  \cdots       & \cdots & \cdots   &  K_{N-2}  & -K_{N-2}  -K_{N-1}  \\
\end{bmatrix} 
\end{align}
This  matrix describes a chain of $N-2$ particles, connected to two fixed particles at $i=1$ and $i=N$ with the springs $K_1$ and $K_{N-1}$ respectively. The eigenmodes of this matrix, $|V_n^{(0)}\rangle$, when padded with zeros at both ends to make them $N$-dimensional, are also eigenmodes of $A$, to zeroth order. Note that $\widetilde{A}$ is identical to $A$ of Eq.~\eqref{Eq.DPT0}, with the difference that it involves $N-2$ particles instead of $N$, and that the first and last springs are $K_1$ and $K_{N-1}$ instead of $k$.

\begin{figure}[t!]
	\includegraphics[width=14cm, keepaspectratio]{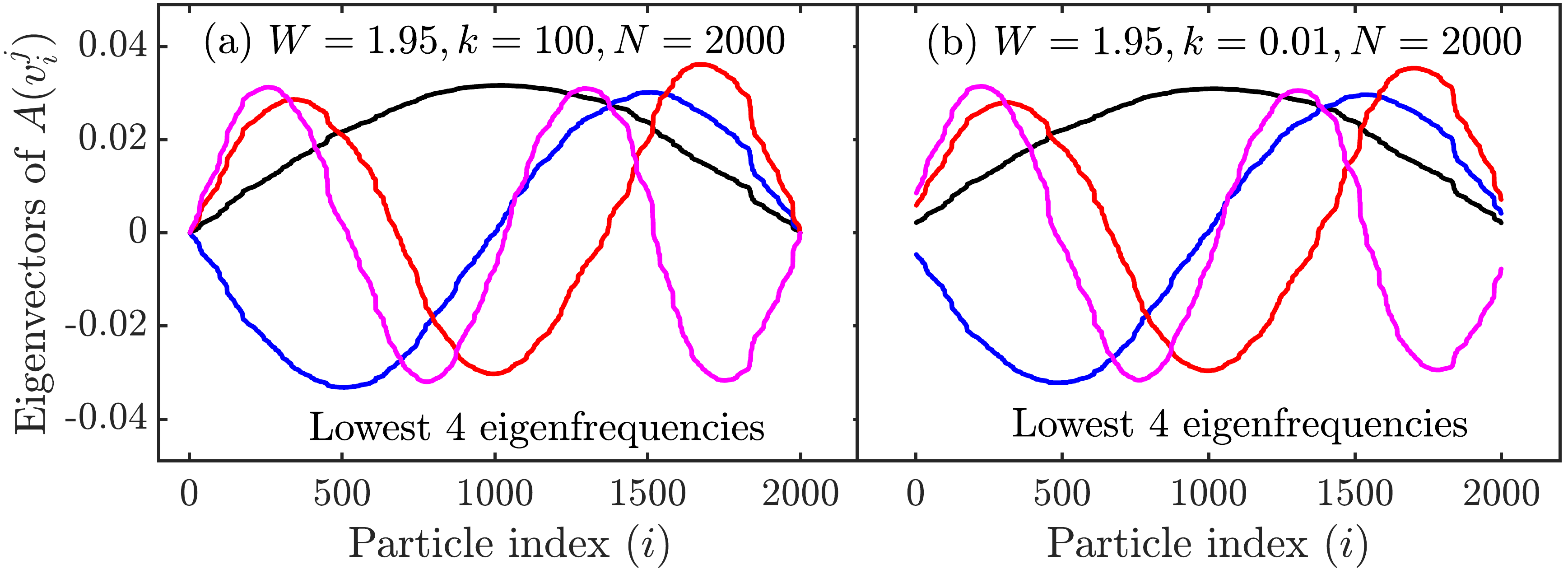}
        \caption{
Eigenvectors associated with the lowest four eigenfrequencies $(j=1,2,3,$ and $4)$ of the matrix $A$ ( see Eq.~\eqref{Eq.DPT0} ) with $N=2000, W=0.5$ and (a) $k=100$ (strong coupling) and (b) $k=0.01$ (weak coupling). In the figure, $v_i^j$ represents vibration amplitude of particle $i$ in the eigenmode $j$. The nature of the eigenvectors at the boundary $(v_1^j$ and $v_N^j)$ are different for $k=100$ and $k=0.01$. For $k=100, |v_1^j|, |v_N^j| \to 0$ for the low-frequency $(\omega \to 0)$ eigenmodes.
}
	\label{Fig:Figev0} 
\end{figure}

Our goal is to understand how $\Braket{1|V_n}$ and $\braket{N|V_n}$ scale with $N$ and $\omega$. In the present context, degenerate perturbation theory~\cite{Sakurai} tells us that
\begin{align}
\Braket{1|V_n^{(1)}}\approx\sum_i
\frac{\Braket{V_i^{(0)}| A^d | V_n^{(0)}}}{\lambda^{(0)}_n-\lambda^{(0)}_i}\Braket{1|V_i^{(0)}}
\end{align}
where the summation is performed only on $\Ket{V_i}$ outside the degenerate eigenspace of $\Ket{V_n}$. In our case there are exactly two modes outside the degenerate subspace -- the boundary modes $\ket{1}$ and $\ket{N}$ -- and consequently the amplitude of the first particle of the $n$-th eigenmode $\Ket{V_n}$ is, to first order,
\begin{align}
\Braket{1|V_n}\approx
\Braket{1|V_n^{(1)}}&=
\frac{\Braket{1|  A^d | V_n^{(0)}}}{\lambda_n^{(0)}+k}
\approx \frac{K_1}{k} \Braket{2|V_N^{(0)}}\ .
\label{Eq:first_order_correction}
\end{align}
That is, to first order, the vibration amplitude of the first particle is proportional to the zeroth order vibration amplitude of the second particle.

In an ordered chain, when all $K_i$ are equal, $\Ket{V_n^{(0)}}$ is a sinusoidal mode which vanishes at $i=1$ and $i=N$, 
\begin{align}
\braket{i|V_n^{(0)}}& \propto \frac{1}{\sqrt{N}}\sin\left(n\pi \frac{i-1}{N-1}\right)\ , 
\end{align}
Therefore, for large $N$ we have $\braket{2|V_n}\propto n$ and since for small $\omega$ dispersion is linear, $n$ is proportional to $\omega$. Combining this with Eq.~\eqref{Eq:first_order_correction}, we get 
\begin{align}
\left\vert \Braket{1|V_n} \right\vert  \approx
\left\vert \Braket{1|V_n^{(1)}} \right\vert  \sim \frac{\omega}{k\sqrt{N}} \ \quad \mbox{(to first order)},
\label{Eq:boundary_scaling} 
\end{align} 
as utilized after Eq.~\eqref{Eq.Lwdth}. An identical argument shows that the vibration amplitude $\braket{N|V_n}$ of the last particle follows the same scaling.

\begin{figure}[t!]
	\includegraphics[width=12cm, keepaspectratio]{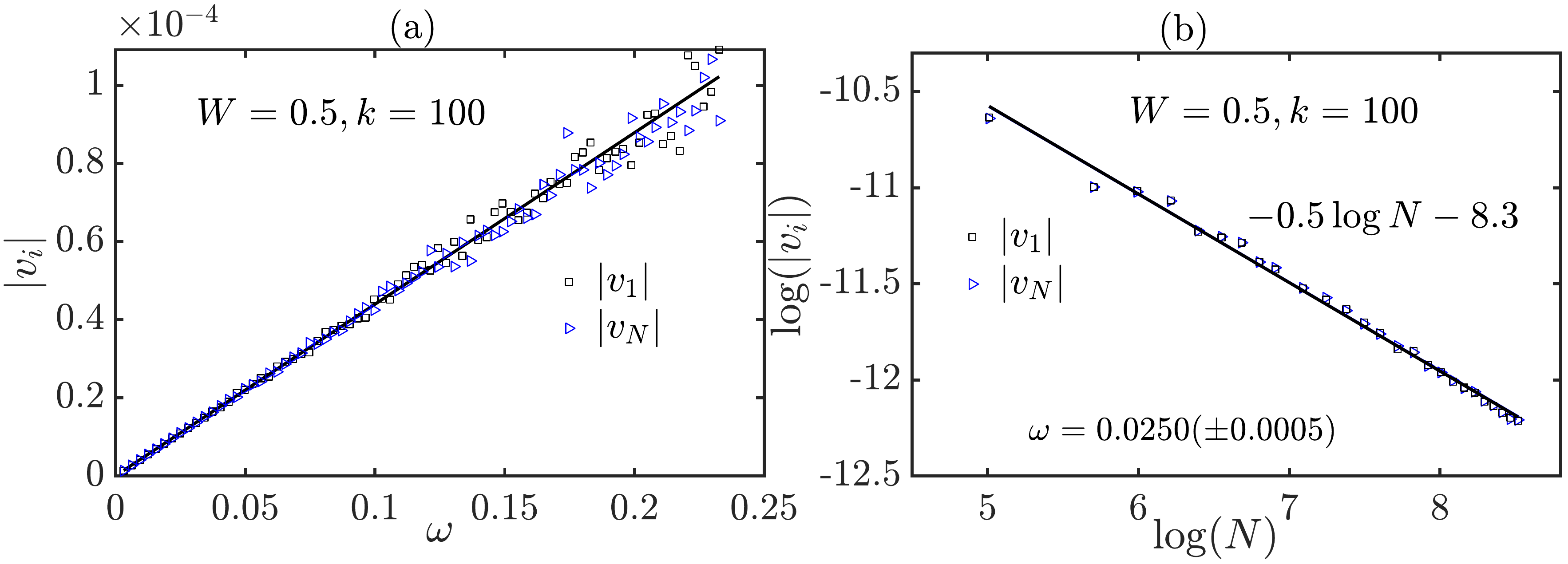}
        \caption{ (a) Dependence of the magnitude of the amplitudes $v_1$ and $v_N$ at sites $1$ and $N$, respectively, on the eigenfrequency $\omega$ for a fixed $N$. Here, we consider a weakly disordered chain (uniform disorder) with $N=2000, W=0.5$ and $k=100$. Results are averaged over $50$ independent realizations of disorder. Different points represent the numerical data while the solid line represents a linear fit to the data for $v_1$. For a given $N$, both $|v_1|$ and $|v_N|$ increase linearly with $\omega$ for low frequencies. (b) For a fixed frequency $\omega$, dependence of $|v_1|$ and $|v_N|$ on the system size $N$. For a given low frequency mode $(\omega = 0.025 \pm 0.0005), |v_1| (|v_N|) \sim \frac{1}{\sqrt{N}}$. Here, we consider $50$ independent realizations of disorder and all the modes having frequencies $\omega \pm 0.0005$. Also,  $W=0.5$ and $k=100$, as in panel (a).
}
	\label{Fig:Figev} 
\end{figure}

Lastly, we argue that Eq.~\eqref{Eq:boundary_scaling} holds for disordered chains as the low-frequency modes are effectively delocalized and follow approximately the same dispersion relation as the ordered chain. This is demonstrated in Fig.~\ref{Fig:Figev0}(a) where we plot the eigenvectors associated with the lowest four eigenfrequencies of a disordered chain with $N=2000, W=1.95$ and $k=100$. Panel (a) shows that the eigenmodes indeed correspond approximately to those of an ordered chain with fixed boundaries. We also verify that the scaling of $|v_1|$ and $|v_N|$ with $\omega$ and $N$ follow Eq.~\eqref{Eq:boundary_scaling} as shown in  Fig.~\ref{Fig:Figev}(a-b). Note that the linear $\omega$ dependence  of $|v_1|$ and $|v_N|$ holds only for the low-frequency modes and fails for $\omega \gg \omega_L$ as the high frequency modes get strongly affected by disorder and become localized (i.e., does not follow the dispersion relation of an ordered harmonic chain).

It is interesting to note that the scaling relation~\eqref{Eq:boundary_scaling}, which holds for $k \gg 1$, is very different from what one would have in case of the weak coupling $(k \ll 1)$. In that case the roles of $A^b$ and $A^d$ in Eq.~\eqref{Eq.DPT0} are interchanged since $A^b$ is much smaller and is treated as a perturbation to $A^d$. Note that $A^d$ describes to a chain whose end particles are free (i.e., not connected to ``external'' springs like $K_1$ and $K_{N-1}$ in $\tilde{A}$), as can be seen from the fact that all rows and columns sum to zero. That is, a global uniform translation is an eigenmode (actually, a Goldstone mode) of the system. In this scenario the modes do not vanish at the boundary but rather their gradient does. This is demonstrated in Fig.~\ref{Fig:Figev0}(b), showing eigenmodes of the same disordered chain as in Fig.~\ref{Fig:Figev0}(a) but with $k=0.01$.

Lastly, we clarify that these notions of ``free'' and ``fixed'' boundary conditions are distinct from the way these notions are used in the literature~\cite{Casher_Lebowitz,Rubin_Greer,DharPhysRevLett2001,ADhar_review}. In their jargon our system always has free boundary conditions, irrespective of the coupling strength, since the two end particles of the disordered chain are not pinned, i.e., these particles are not connected to some external springs other than the ones that couple the system with the heat baths. We only use the notions of ``free'' and ``fixed'' boundary conditions in the context of the eigenmode behavior near the boundaries under weak and strong coupling, respectively.

\section{Density of states}

The density of states (DOS), $\rho(\omega)$, gives the number of states per unit frequency interval at the frequency $\omega$. It is defined as
\begin{equation}
\rho(\omega) =\left\langle \frac{1}{N} \sum_{\omega_q} \delta (\omega_q - \omega)\right\rangle,
\end{equation}
where $N$ is the number of particles in the disordered chain and $\left\langle \right\rangle$ indicates averaging with respect to the independent realizations of the disorder. Thus, we have $\int {\rm d} \omega \rho(\omega) = 1$.
We compute $\rho(\omega)$ for the isolated disordered chain by constructing the histogram of the frequencies $\omega(= \sqrt{\lambda})$ obtained by diagonalizing the following $N \times N$ matrix:
\begin{align}
H=
\begin{bmatrix}
K_1      &  -K_1    & 0        & \cdots     & 0 \\
-K_1     &  K_1+K_2 & -K_2     & \cdots     & 0   \\
0        &  -K2     & K_2+K_3  & \cdots     & 0   \\
0        & \cdots   & \cdots   & \cdots     & -K_{N-1}  \\
0        &  0       & \cdots   & \cdots     & K_{N-1}
\end{bmatrix}
\label{Eq:Diagnl}
\end{align}
where $\lambda$ are the eigenvalues of $H$. For an ordered chain, we know~\cite{Ashcroft}
\begin{equation}
\rho(\omega) =\frac{2}{\pi \sqrt{\omega_{m}^2-\omega^2}},
\label{Eq.rho_order}
\end{equation}
where $\omega_{m}=2\omega_0$ is the maximum allowed frequency for the phonons, and $\omega_0$ is the natural frequency of an ordered chain. Thus, for an ordered chain $\rho(\omega)$ diverges as $\omega \rightarrow \omega_{m}$, as discussed in the main text and consequently $\rho(\omega)=0$ for $\omega>2\omega_0$. The divergence of density of states at $\omega=\omega_{m}$ for the ordered chain $(W=0)$ corresponds to the van-Hove singularity~\cite{Ashcroft}.

\section{Details of the transfer matrix method to compute transmission coefficient, $\tau(\omega)$}\label{Ap:TMM}

To compute the transmission coefficient $\tau(\omega)$ for a phonon of frequency of $\omega$ as it passes through the disordered chain of length $L$, we construct the transfer matrix for the disordered chain as described below.
We can recast Eq.~\eqref{Eq:EOM2} in the following matrix form:
\begin{align}
\begin{pmatrix}
x_{n+1}\\
x_{n}
\end{pmatrix}
&=
\begin{pmatrix}
\frac{K_{n}+K_{n-1}-m_n\omega^2}{K_{n}} & - \frac{K_{n-1}}{K_{n}}\\
1            & 0
\end{pmatrix}
\begin{pmatrix}
x_{n}\\
x_{n-1}
\end{pmatrix}\nonumber \\
&=M_n
\begin{pmatrix}
x_{n}\\
x_{n-1}
\end{pmatrix},
\label{Eq.EOMe}
\end{align}
where, $m_n = M $ for $(n=1,2, \cdots, N)$ and $m_n=m$ otherwise.
We can relate the two ends of the disordered chain using the matrix $M_n$ iteratively, and thus we have
\begin{align}
\begin{pmatrix}
x_{N+2}\\
x_{N+1}
\end{pmatrix}
=\prod_{n=0}^{N+1}M_n
\begin{pmatrix}
x_{0}\\
x_{-1}
\end{pmatrix}
\label{Eq.EOM4}
\end{align}
Eq.~\eqref{Eq.EOM4} relates the amplitudes of the waves on the two sides (i.e. in the ordered region) of the disordered chain. In the ordered region, we can express the amplitude of the displacement at any lattice site as the superposition of plane waves. For example, in the region $n>N$, for phonons with wave number $q$, we can write:
\begin{align}
x_n=A\mathrm{e}^{i\chi n} + B \mathrm{e}^{-i\chi n},
\end{align}
where, $A$ and $B$ are two constants. The left hand side of Eq.~\eqref{Eq.EOM4} can be written as
\begin{align}
\begin{pmatrix}
x_{N+2}\\
x_{N+1}
\end{pmatrix}
&= 
\begin{pmatrix}
1  &   1\\
\mathrm{e}^{-i\chi} & \mathrm{e}^{i\chi}
\end{pmatrix}
\begin{pmatrix}
A \mathrm{e}^{i\chi (N+2)} \\
B \mathrm{e}^{-i\chi (N+2)}
\end{pmatrix}
\nonumber \\
&=
Q
\begin{pmatrix}
A \mathrm{e}^{i\chi (N+2)} \\
B \mathrm{e}^{-i\chi (N+2)}
\end{pmatrix}
\label{Eq.EOM5}
\end{align}
Now, multiplying both sides of Eq.~\eqref{Eq.EOM4} by $Q^{-1}$ from the left, we get
\begin{align}
Q^{-1}
\begin{pmatrix}
x_{N+2}\\
x_{N+1}
\end{pmatrix}
&=
Q^{-1} \left[\prod_{n=0}^{N+1}M_n\right]QQ^{-1}
\begin{pmatrix}
x_{0}\\
x_{-1}
\end{pmatrix}
\nonumber \\
&= T^{(N)}
Q^{-1}
\begin{pmatrix}
x_{0}\\
x_{-1}
\end{pmatrix}
\label{Eq.EOM6}
\end{align}
where, 
\begin{align}
Q^{-1}=\frac{1}{2i\sin \chi}
\begin{pmatrix}
\mathrm{e}^{i\chi}   & -1\\
-\mathrm{e}^{-i\chi} &  1
\end{pmatrix}
\label{Eq.EOM7}
\end{align}
and we have introduced the transfer matrix $T^{(N)}$, connecting the waves on two sides of the disordered chain, as
\begin{align}
T^{(N)}=Q^{-1} \left[\prod_{n=0}^{N+1}M_n\right]Q
\label{Eq.TMF}
\end{align}
\begin{figure}[t!]
\includegraphics[width=12cm,keepaspectratio]{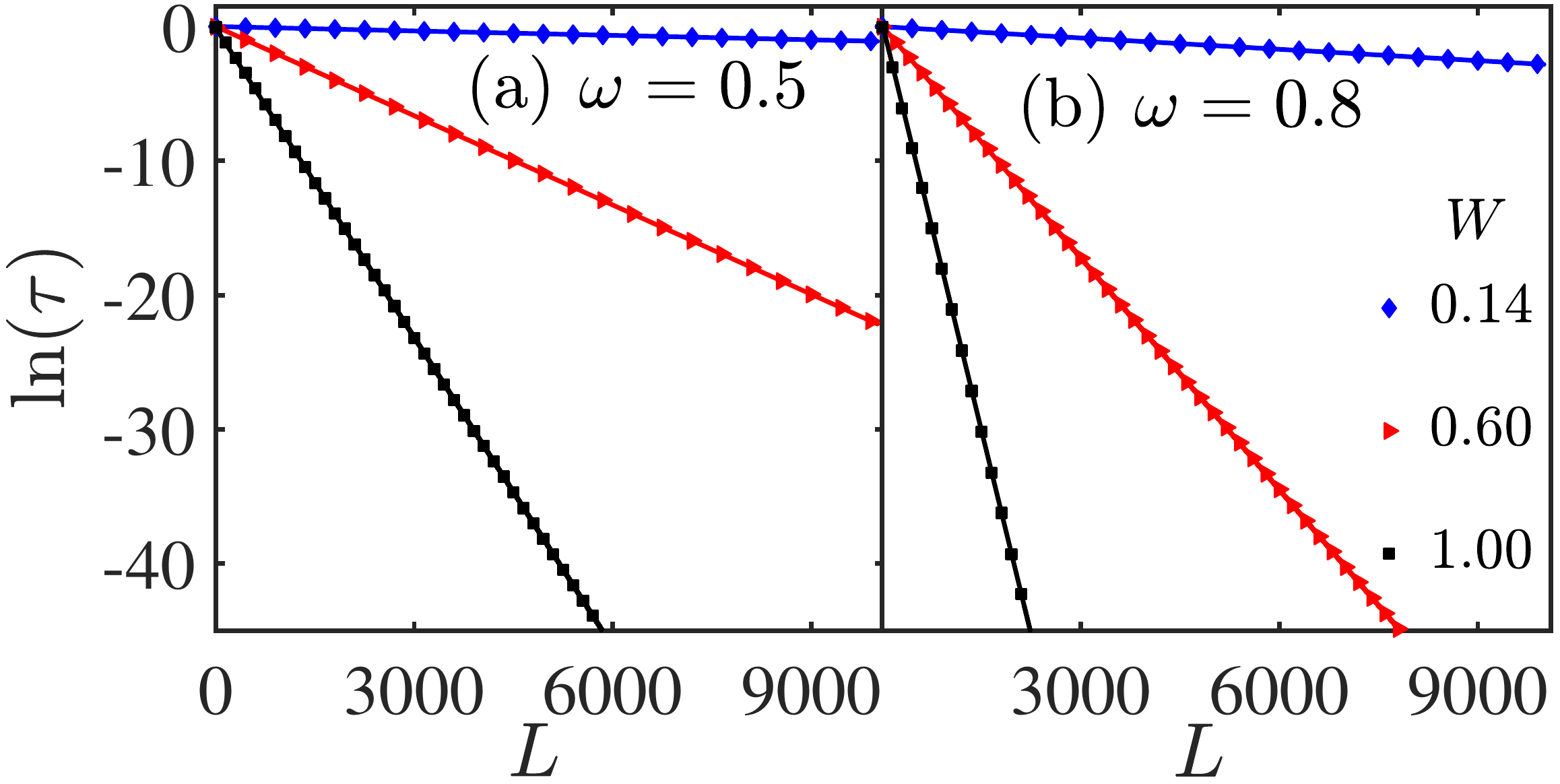}
\caption{ Dependence of $\ln \tau (\omega)$ on the system size $(L)$ of the uniformly disordered chain (model 1) for different strength of disorder $W$ at a given frequency $\omega$:  (a) $\omega=0.5$ and (b) $\omega=0.8$. Thick dots represent the actual data while the solid lines are the liner fit $[ \ln \tau (\omega) = -L/\xi(\omega) ]$ to the respective numerical data. For a given value of $W$, $ \ln \tau (\omega)$ decays linearly with $L$ for any $\omega$ and the slope of the curve gives the localization length $\xi(\omega)$. Results are averaged over $200$ independent realizations of disorder.
}
\label{Fig:LnTR2}
\end{figure}
Thus, from Eq.~\eqref{Eq.EOM6}, we have
\begin{align}
\begin{pmatrix}
 {e}^{i\chi}  x_{N+2} - x_{N+1}\\
-{e}^{-i\chi} x_{N+2} + x_{N+1}
\end{pmatrix}
&=T^{(N)}
\begin{pmatrix}
 {e}^{i\chi}  x_{0} - x_{-1}\\
-{e}^{-i\chi} x_{0} + x_{-1}
\end{pmatrix}
\label{Eq.EOM8}
\end{align}
We can simplify Eq.~\eqref{Eq.EOM8} further by considering the following situation. Let us assume a phonon with wave number by $q$ coming from the right side $(n > N)$ of the disordered chain. There is only a transmitted wave on the left side $(n < 1)$ of the disordered chain while on the right side of the system we can have a reflected wave along with the incident wave. Therefore the amplitudes $x_0$ and $x_{-1}$ of the wave with frequency $\omega$ can be taken as 
\begin{align}
x_0=1, x_{-1} = \mathrm{e}^{i\chi}.
\label{Eq.init}
\end{align}
Here, wave number $q$ and frequency $\omega$ are related by the phonon dispersion relation~\cite{Ashcroft}
\begin{equation}
\omega^2(q)=\frac{4k}{m}\sin^2\left(\frac{qa}{2}\right)=\frac{4k}{m}\sin^2\left(\frac{\chi}{2}\right)
\label{Eq.Disp}
\end{equation}
From Eq.~\eqref{Eq.init}, we obtain
\begin{align}
 {e}^{i\chi}  x_{0} - x_{-1}&= 0\\
-{e}^{-i\chi} x_{0} + x_{-1}&= 2i \sin \chi.
\label{Eq_EOM8}
\end{align}
Thus, Eq.~\eqref{Eq.EOM8} reduces to
\begin{align}
\begin{pmatrix}
 {e}^{i\chi}  x_{N+2} - x_{N+1}\\
-{e}^{-i\chi} x_{N+2} + x_{N+1}
\end{pmatrix}
&= 2i \sin \chi
\begin{pmatrix}
T^{(N)}_{12}\\
T^{(N)}_{22}
\end{pmatrix}
\label{Eq.EOM9}
\end{align}
where $T^{(N)}_{ij}$ represents the $(i,j)$ element of the transfer matrix $T^{(N)}$ defined in Eq.~\eqref{Eq.TMF}. Finally, the transmission coefficient, $\tau(\omega)$, for an incoming wave with frequency, $\omega$, is given by~\cite{Soukoulis}
\begin{align}
\tau(\omega)=\frac{1}{\vert T_{22}^{(N)}\vert^2}= \frac{4\sin^2 \chi}{\vert {e}^{-i\chi} x_{N+2} - x_{N+1} \vert^2}.
\label{Eq.EOM10}
\end{align}
Since, in our case the disorder does not break the time reversal symmetry, the transmission coefficient for phonons coming from the right side of the disordered system is same as those coming from the left side of the system~\cite{Soukoulis}.
\begin{figure}[t!]
	\includegraphics[width=12cm, keepaspectratio]{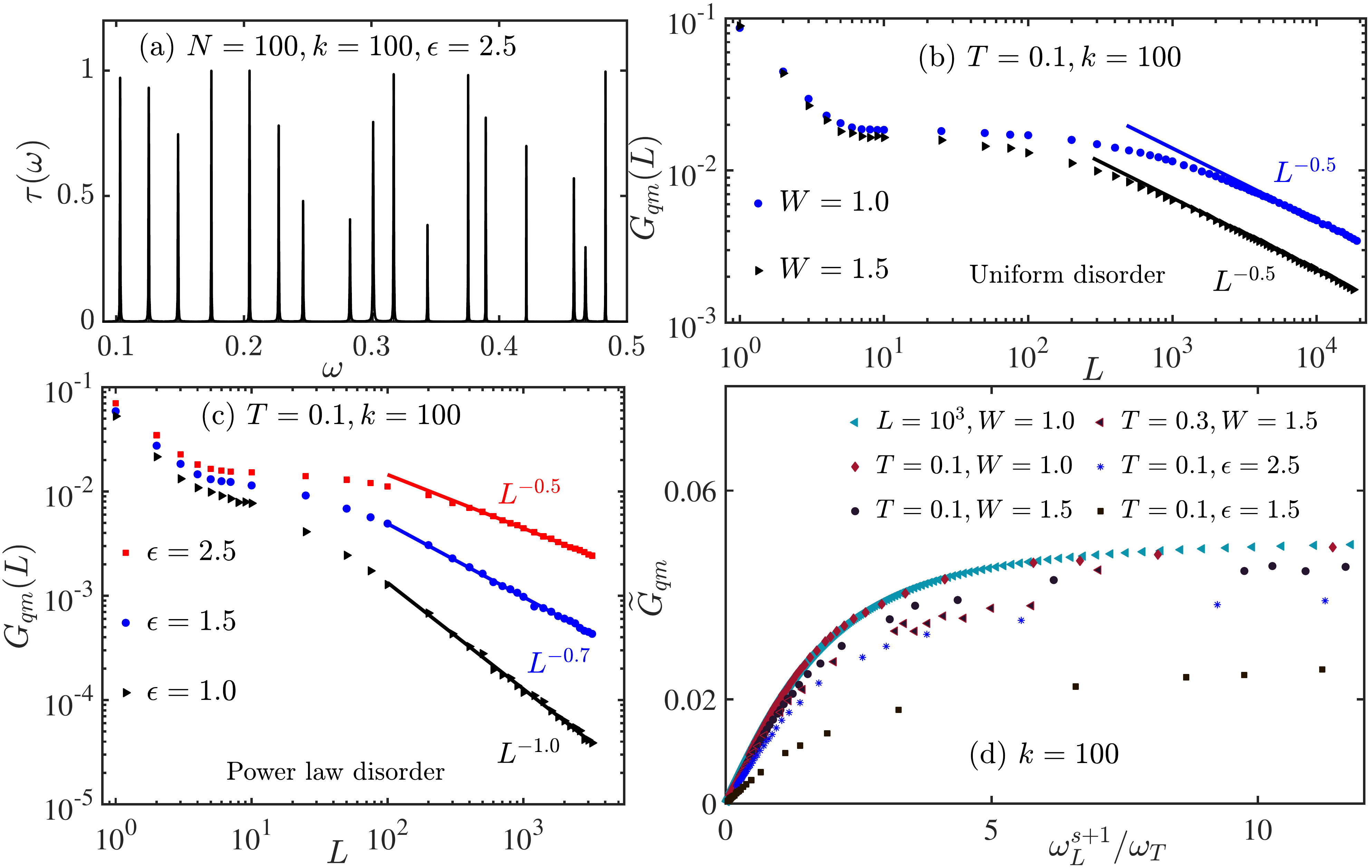}
        \caption{ (a) Transmission coefficient $\tau(\omega)$ for power law disorder with disorder strength $\epsilon=2.50, k=100,$ and $N=100$. The black dashed vertical lines indicate the eigenfrequencies of the disordered chain, and their height corresponds to Eq.~\eqref{Eq:TEq2}. The dependence of the thermal conductance $G_{qm}(L)$ on the system size $(L)$ for different disorder strengths: (b) for uniform disorder and (c) for a power law distribution of the disorder. We set  $k=100$ and $T=0.10$. In panel (b-c), thick dots represent the actual data while the solid line shows the power-law fit to the original data for large $L$. The scaling exponents for different disorder strengths remain unaffected for $k=100$. (d) Thermal conductance for one dimensional disordered harmonic chains with different models of disorder in terms of the variable $\omega_L^{s+1}/\omega_T$ (see main article for details). All data collapse approximately for small $\omega_L^{s+1}/\omega_T$ (.i.e., large $L$), as expected. 
}
	\label{Fig:largeKb}
\end{figure}

Using Eq.~\eqref{Eq.EOM4}, one can obtain $x_{N+2}$ and $x_{N+1}$ and then calculate the transmission coefficient, $\tau(\omega)$, using Eq.~\eqref{Eq.EOM10}.
For a given $L$, once we obtain the transmission coefficient for different $\omega$, we can compute the thermal conductance for the disordered chain using the Landauer formula given in main article. Finally, we average the results over $500$ independent realizations of the disorder.

\section{Dependence of transmission coefficient on $L$ for a fixed frequency}

For a given strength of disorder, the transmission coefficient $\tau(\omega)$ depends on the length $L$ of the disordered chain. The $L$-dependence of $\ln \tau(\omega)$ for two arbitrarily chosen frequencies: (a) $\omega=0.5$ and (b) $\omega=0.8$ are shown in Fig.~\ref{Fig:LnTR2}(a-b). We set $k=1$ and $m=1$. In general, we find that $\langle \ln \tau(\omega)\rangle$ decays linearly with $L: \langle \ln \tau(\omega)\rangle \propto {-\frac{L}{\xi(\omega)}}$ for different different values of $W$. Thus, for a given $W$, the inverse of the slope of $\langle \ln \tau(\omega)\rangle- L$ curve gives us the localization length associated with the particular mode of frequency $\omega$.

\section{Scaling of thermal conductance with $L$ for strong coupling, $k (\gg 1)$}

In the main article, we discuss results for $k=0.01$ (representing weak coupling between the system and the heat bath) and  $k=1.0$ (intermediate coupling). Here, we show the effect of large $k (\gg1)$ on the transmission coefficient and the scaling of thermal conductance. The transmission coefficient for different frequencies with $k =100$ is shown in Fig.~\ref{Fig:largeKb}(a), where we consider a disordered chain (power law disorder) with $N=100$ and disorder strength $\epsilon=2.50$. For $k=100$, $\tau(\omega)$ develops well separated Lorentzians having area and width given by Eq.~\eqref{Eq.Larea}-\ref{Eq.Lwdth}. We have also verified that the scaling exponents for thermal conductance with $k=100$ and $k=0.01$ remain same in different disorder regimes as demonstrated in Fig.~\ref{Fig:largeKb}(b) (for uniformly disordered chain) and Fig.~\ref{Fig:largeKb}(c) (for power-law disorder). When expressed in terms of the variable $\omega_L^{s+1}/\omega_T$, thermal conductance for one dimensional disordered harmonic chains with different models of disorder collapse approximately for small $\omega_L^{s+1}/\omega_T$~(see Fig.~\ref{Fig:largeKb}(d)). For a given $T$, we expect this collapse to work for large $L$ (.i.e., small $\omega_L$ and fixed $\omega_T$). 

\section{Scaling of thermal conductance with $T$ for different coupling $k$}
\begin{figure}[t!]
	\includegraphics[width=12cm, keepaspectratio]{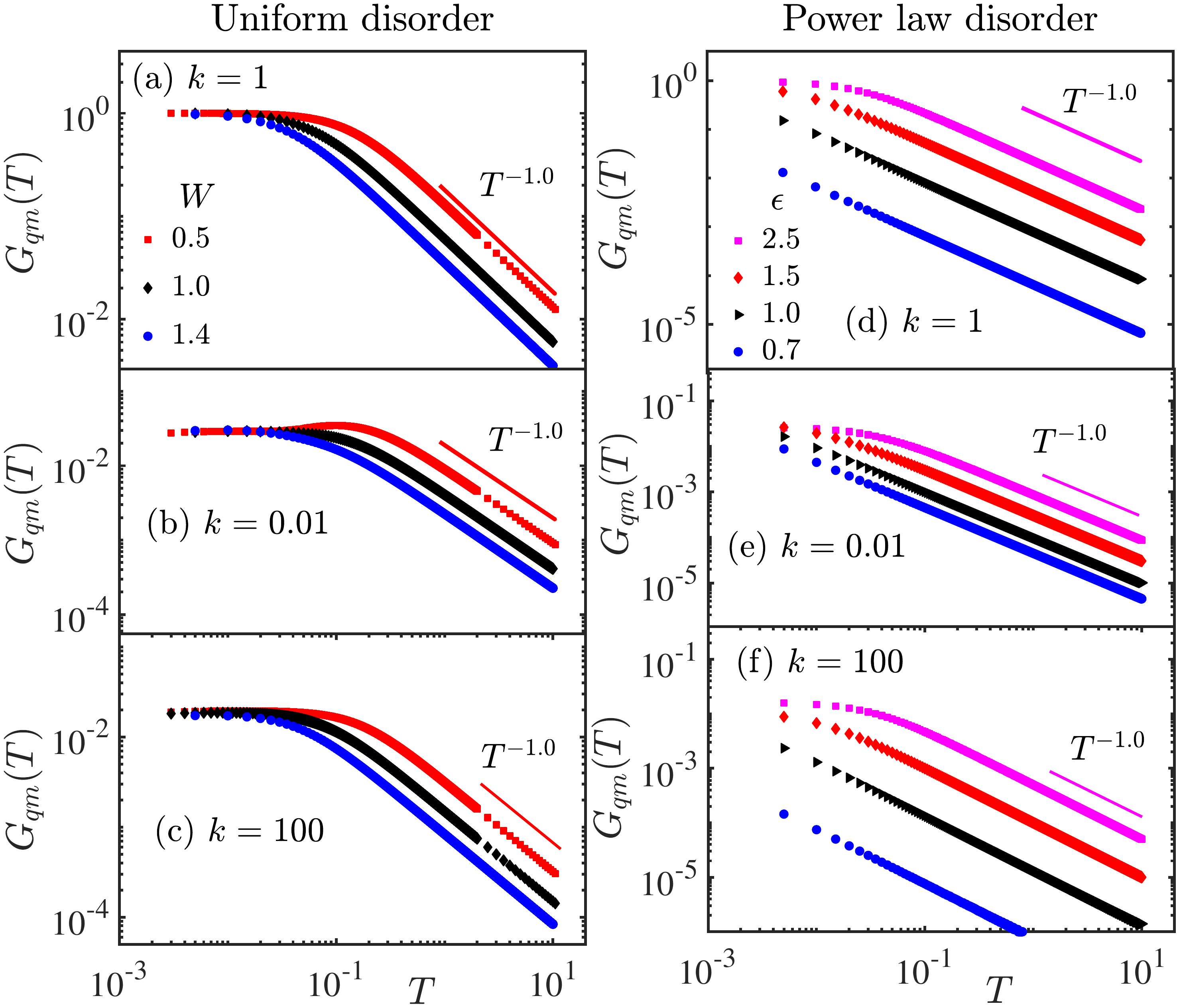}
        \caption{ The $T$-dependence of the thermal conductance, $G_{qm}(T)$, for a fixed length $L(=1000)$ of the disordered chain for uniform disorder ((a) $k=1.0$, (b) $k=0.01$ and (c) $k=100$) and power law disorder ((d) $k=1.0$, (e) $k=0.01$ and (f) $k=100$). For large $T (>\hbar\omega_L), G_{qm}(T) \propto T^{-1}$ irrespective of the disorder and coupling strength $(k)$.
}
	\label{Fig:TKb}
\end{figure}

The dependence of thermal conductance $G_{qm}(T)$ on temperature $T$ for a given $L$ for different disorder and coupling strength $(k)$  is shown in Fig.~\ref{Fig:TKb}(a-f). We first analyze the case $k=1.0$. We find that $G_{qm}(T)$ is close to unity for small $T (\ll \hbar\omega_L)$ and it decays as $T^{-1}$ for $T > \hbar\omega_L$ in all disorder regimes. Here, $\xi(\omega_L)=L$. For a given $L$, we can understand the dependence of $G_{qm}(T)$ on $T$ by considering the following observations. Phonons with frequencies such that the associated localization length is less than the system size, $L$, are localized and consequently do not contribute to the thermal conductance. If $T$ is such that $\omega_T< \omega_L$ then the modes which are delocalized are thermally excited. For a given $L$, as $T$ is increased from very small value to $T^*_W = \hbar \omega_L$, more and more delocalized (ballistic) phonons contribute to the thermal conductance $G(L,T)$ and consequently $G(T)$ should initially increase with $T$ for a given $L$. In Fig.~\ref{Fig:TKb}(a), we find that for a given $W$, $G_{qm}(T)$ remains unity for $T \ll T^*_W$ implying that $G(T)$ increases \textit{linearly} with $T$ for $T \ll T^*_W$ (note that $T$ is included in the definition of $G_{qm}(L,T)$). For $T>T^*_W$, the modes which are localized $(\xi(\omega)<L)$ are thermally excited and these modes do not contribute to the heat transport, causing $G(T)$ to saturate for $T > T^*_W$. Such a saturation is reflected as the linear decay of $G_{qm}(T) (\propto T^{-1})$ in Fig.~\ref{Fig:TKb}(a), which is independent of the disorder strength, $W$. Note that for a given $L, T^*_W$ decreases with increasing $W$ and subsequently the temperature regime for $G_{qm}\approx 1$ shrinks to smaller values with increasing $W$.  

We also find that the scaling $G_{qm}(T) \propto T^{-1}$ for $T>\hbar\omega_L$ holds irrespective of nature of disorder and coupling strength (see Fig.~\ref{Fig:TKb}(a-f)), following the theoretical prediction made in the main article.

\end{document}